\documentclass[conference]{IEEEtran}
\IEEEoverridecommandlockouts
\usepackage{cite}
\usepackage{amsmath,amssymb,amsfonts}
\usepackage{algorithmic}
\usepackage{graphicx}
\usepackage{textcomp}
\usepackage{xcolor}
\usepackage{float}
\usepackage{siunitx}
\usepackage{multirow}
\usepackage{booktabs}
\usepackage{commath}
\usepackage{relsize}
\usepackage{amssymb}

\usepackage[linesnumbered,ruled,vlined]{algorithm2e}
\usepackage{etoolbox}
\usepackage{todonotes}
\usepackage{cuted}
\usepackage{url}
\newcounter{algoline}
\usepackage{caption}
\usepackage{subcaption}
\usepackage{textcomp}

\usepackage{siunitx}
\newcommand\Numberline{\refstepcounter{algoline}\nlset{\thealgoline}}
\AtBeginEnvironment{algorithm}{\setcounter{algoline}{0}}

\newcommand{\etal}{et al.}
\newcommand{\figureref}{Figure \ref}
\newcommand{\sectionref}{Section \ref}
\newcommand{\tabref}{Table \ref}
\newcommand{\tmin}{\textminus}
\newcommand{\ttimes}{$\times$}

\newcommand{\SubItem}[1]{
{\setlength\itemindent{15pt} \item[-] #1}
}

\SetKwInput{kwInit}{Init}
\SetKwInput{kwInput}{Input}
\SetKwInput{kwOutput}{Output}
\graphicspath{{pfigures/eps/}{pfigures/pdf/}{pfigures/png/}}

\def\BibTeX{{\rm B\kern-.05em{\sc i\kern-.025em b}\kern-.08em
    T\kern-.1667em\lower.7ex\hbox{E}\kern-.125emX}}

\begin{document}

\title{Generation of Realistic Cloud Access Times for Mobile Application Testing using Transfer Learning\\
\thanks{This work has been partially supported by the German Federal Ministry of Education and Research (BMBF) within the project SecureFog under the grant number 16KIS0777.}
}

\author{\IEEEauthorblockN{1\textsuperscript{st} Manoj R. Rege}
\IEEEauthorblockA{\textit{Technische Universität Berlin} \\
Berlin, Germany\\
manojrege@mailbox.tu-berlin.de}
\and
\IEEEauthorblockN{2\textsuperscript{nd} Vlado Handziski}
\IEEEauthorblockA{\textit{R3 Reliable Realtime Radio} \\
Berlin, Germany \\
vlado.handziski@r3coms.com}
\and
\IEEEauthorblockN{3\textsuperscript{rd} Adam Wolisz}
\IEEEauthorblockA{\textit{Technische Universität Berlin} \\
Berlin, Germany\\
wolisz@tkn.tu-berlin.de}}

\maketitle

\begin{abstract}
    The network Quality of Service (QoS) metrics such as the access time, the bandwidth, and the packet loss play an important
    role in determining the Quality of Experience (QoE) of mobile applications. Various factors like the Radio Resource
    Control (RRC) states, the Mobile Network Operator (MNO) specific retransmission configurations, handovers triggered
    by the user mobility, the network load, etc. can cause high variability in these QoS metrics on 4G/LTE, and WiFi
    networks, which can be detrimental to the application QoE. Therefore, exposing the mobile application to realistic network
    QoS metrics is critical for a tester attempting to predict its QoE. A viable approach is testing using synthetic traces.
    The main challenge in the generation of realistic synthetic traces is the diversity of environments and the lack of wide scope of
    real traces to calibrate the generators. In this paper, we describe a measurement-driven methodology based on
    transfer learning with Long Short Term Memory (LSTM) neural nets to solve this problem. The methodology requires a
    relatively short sample of the targeted environment to adapt the presented basic model to new environments, thus
    simplifying synthetic traces generation. We present this feature for realistic WiFi and LTE cloud access time models
    adapted for diverse target environments with a trace size of just 6000 samples measured over a few tens of minutes.
    We demonstrate that synthetic traces generated from these models are capable of accurately reproducing application QoE
    metric distributions including their outlier values.
\end{abstract}


\section{Introduction}
\label{sec:intro}
The Quality of Experience (QoE) of mobile applications is highly dependent on factors such as the access network
Quality of Service (QoS), and the user context, among others. It has been observed that in the widely deployed
cellular networks like 3G, 4G/LTE, and WiFi, various problems in the network stack can cause significant variability
in the network QoS metrics such as the access time latency, bandwidth, packet loss, etc. Such high variability can have
an adverse impact on the mobile application
QoE~\cite{10.1145/2307636.2307658,10.1145/2639108.2639115,10.1145/2535372.2535399,DBLP:conf/pam/XuNM19,Shafiq:2013:FLC:2494232.2465754,Huang:2010:AAP:1814433.1814452,10.1145/2906388.2906393,DBLP:conf/imc/GemberAPVC12}.
For example, when using a cellular network like LTE, Radio Resource Control (RRC) radio link layer control states which
are used by the base station to coordinate with the device have a significant impact on application performance and
power consumption~\cite{10.1145/2307636.2307658,10.1145/2639108.2639115}. On
the data plane, Mobile Network Operators (MNOs)
often employ protocols configurations in the Radio Link Control (RLC) that can adversely impact the performance of
transport protocols like TCP, thus degrading the application
QoE~\cite{10.1145/2535372.2535399}. External user context
factors like mobility lead to handover within the network that can introduce large access time latencies and degrade
application throughput~\cite{DBLP:conf/pam/XuNM19}. When a large number of cell
users are connected to the network,
there can be RRC failures for new users either blocking their network access or increasing the connection establishment
latency~\cite{Shafiq:2013:FLC:2494232.2465754}. At certain times of the day,
overall higher demand for video-streaming,
web applications can increase the aggregate network traffic volume in the downlink, thereby degrading the throughput and
the access time latency which is detrimental to application
QoE~\cite{Huang:2010:AAP:1814433.1814452}.
With the wide-scale roll-out of the 5G networks, some of these problems could fade away. Also, the situation might ease out if
the slices supporting the stable network QoS and those matching the individual application requirements will be rolled out.
But the scope of availability of such application-specific slices in the near future is not quite clear.

\begin{figure*}
    \includegraphics{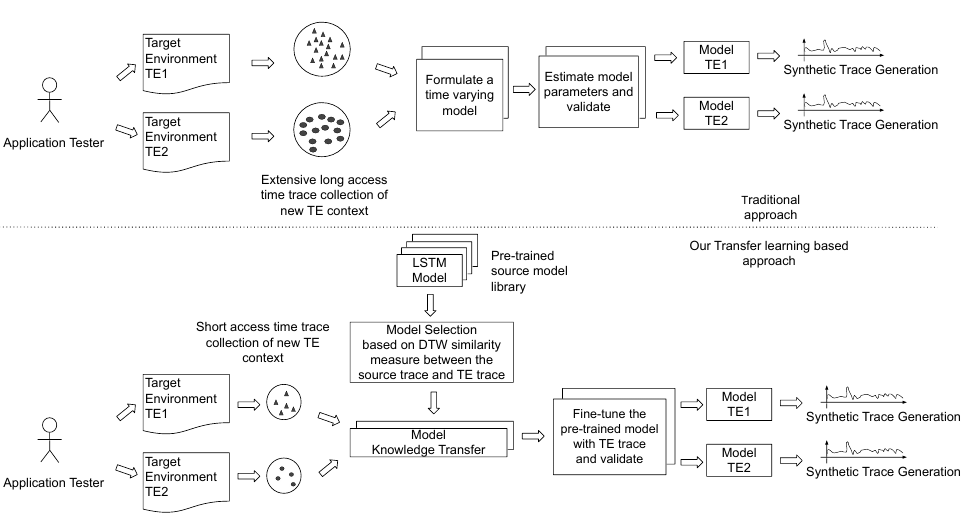}
    \caption{Cloud access time modeling: A comparison of our transfer learning based approach with the traditional approach.\label{fig:approach-comparison}}
\end{figure*}

Continuous measurement of application QoE and its improvement has a strong incentive for the application testers, as it is
essential for the business's success. The application testers require a good understanding of the network QoS delivered to
the users in the real world. This can help them to fine-tune application-specific protocols and parameters to improve their
QoE. For example, in video streaming based applications the buffer sizes and streaming rates need to fine-tuned based on
the network QoS~\cite{8424813}. We argue that there is a common need across the
various application use-cases to study the
impact of network QoS on the application QoE in a given context which is limited to specific scope of interest. For instance,
measuring QoE in a variety of scenarios such as users physical environment viz. indoors, outdoors, in a cafe, or a
university, etc, within a metropolitan region, while the user is mobile; commuting, driving a vehicle, etc., during certain
hours of the day, on a specific MNO, on a specific mobile device model, or any other such factors that impact the network
QoS and thus the application QoE. We refer to such narrow scope of QoE testing as target environment context-driven testing.
In this work, we focus on target environment context-driven testing under the impact of cloud access time network QoS metric
as it has the strongest influence on the perceived overall QoE for numerous
mobile applications. Amazon~\cite{10.1145/1281192.1281295}
estimates that an increase of access time latency by 100~ms can lead to 1\% annual sales loss. A similar study from
Google~\cite{marissa} has found that 500~ms latency per transaction can result
in loss of traffic as high as 20\%.

Unfortunately, the current approach for QoE testing is not suited for carrying out target environment context-driven
testing. It typically makes use of a strongly oversimplified access time model represented by the mean and standard deviation
statistics of the well-known parametric distributions~\cite{testdevblog}. These
statistics are either derived
through common knowledge or from the literature. Such an approach has fundamental shortcomings.
Firstly, the access time statistics obtained from literature,
databases~\cite{opensig}
do not include the temporal dynamics of the access time variations; hence they
lack realism. Secondly, these statistics are often generic, labeled by location
and access network type --- WiFi, 3G, and LTE. They lack context-based labels
related to the trace collection environment such as stationary, mobility, indoors,
outdoors, etc. Joorabchi \etal~\cite{6681334} in their survey-based study of
mobile
application developers and testers found the lack of support to accurately mimic realistic network
environments as one of the top challenges in mobile application development and testing.

According to~\cite{Noble:1997:TMN:263105.263140,browserstack_testing}, there
are three main approaches for carrying out
more rigorous application testing viz. real, emulation, and simulation models.
The real approach includes exposing the application on a real mobile device to real network conditions directly.
While this approach seems to have the highest level of realism, it is hard to cover a large enough scope of conditions.
Furthermore, the real network conditions cannot be controlled and testing is not repeatable. An alternative is to build
a comprehensive trace archive that contains cloud access time traces and replay them to emulate real network conditions
~\cite{DBLP:journals/corr/GoelSWLF17}. While the trace-based emulation approach
is valuable, such open archives are not
available. Further, it is also likely that any such archive would have to be continuously updated given the high
variability of target environments and contexts. The simulation modeling approach includes building a model for generating
synthetic cloud access time traces that mimic a large scope of real-world conditions in a
controllable and statistically repeatable way. Models have the potential to be representative of
the broader population of users and scenarios if they properly interpolate and extrapolate the set
of measurements used for its creation. Unfortunately, models might also need frequent re-calibration
if the environment changes.

Building good models is non-trivial. The traditional modeling approach (shown in  \figureref{fig:approach-comparison})
requires an extensive long collection of traces for a representative set of target environments.
According to~\cite{HIRTH201585}, trace collection for network modeling requires
careful consideration of parameter
values for granularity, duration, scale, diversity of measurement points for traces. Next, the traces need to be
processed and then reduced to parameters of a time-varying
model~\cite{Noble:1997:TMN:263105.263140}. Thereafter, the
parameters of the model need to be estimated and the model validated. For a newer target environment, the entire set of
steps need to be repeated again. Thus, building models is hardly a feasible approach for a tester of a single application or
a small number of applications. Most application testers are not networking experts and lack of sufficient domain
knowledge makes both data collection and building such models challenging for them. Therefore, the rigorous network
modeling approach has been of limited interest within the application testing
community in industry~\cite{7287739}.

This paper aims to lower the barriers to building realistic cloud access time models for the generation of synthetic
traces that can be used in target environment context-driven testing. We develop a
transfer learning based approach to build a model for a target environment (shown in  \figureref{fig:approach-comparison}).
We claim and demonstrate that our approach requires a relatively short measurement trace of cloud access
time within a target environment to build its realistic model as compared to traditional approaches to model realistic
cloud access times~\cite{1007528,4146998} that require extensively long
targeted environment traces.
We build a library of so-called source models by training a Long Short Term Memory (LSTM) neural net
architecture on an extensive set of traces of cloud access times collected over WiFi as well as LTE networks in various
contexts. These contexts include scenarios such as indoor, and outdoor environments, stationary, and mobility in the
train, vehicle, and walking. Then to build the model of a newer target environment, an appropriate
pre-trained model is selected from the library based on the Dynamic Time Warping (DTW) similarity measure between the
source and the target traces. The learned features (the net's weights) of the selected pre-trained model are then
transferred in the lower layers of the selected pre-trained source model by freezing them, also called knowledge
transfer. Finally, the model is fine-tuned by training on the higher layers with the target environment trace and
then validated. The main challenges in applying transfer learning to the trace generation problem are summarized by the
following open questions. (1) Which pre-trained source model to select as the source for the knowledge transfer? (2) How much knowledge to transfer? (3) How much target environment trace will be sufficient
for fine-tuning to generate realistic traces? In this paper, we pursue a systematic investigation to answer
these questions.

The main contributions of our work are as follows:

\begin{itemize}

    \item {Source model generation}
    \SubItem{We propose a generic LSTM neural net architecture for modeling cloud access times that is instantiable
    across any target environment.}
    \SubItem{We have built an access time trace archive spanning across target environments that include user mobility (vehicle, train, walk), a stationary user at different locations (cafe, dormitory,
    university campus, home, office) on WiFi and LTE networks.}
    \SubItem{Through systematic training and validation,
    we instantiate the LSTM architecture for each of these environment traces in the archive, thus resulting in a library of 10 pre-trained LSTM source models. This library of source models can be extended as needs appear.}

    \item {Target model development}
    \SubItem{Our simplified library of pre-trained source models covering a spectrum of useful environments can be directly used for obtaining models of the newer target environments through fine-tuning on its short sample trace.
    This eliminates the need for tedious long trace collection.}
    \SubItem{We suggest how to select the source model for a
    given target environment using the similarity measure like Dynamic Time Warping (DTW) and demonstrate the usefulness of such selection.}
    \SubItem{Using the Symmetrical Mean Absolute Percentage Error (SMAPE) as the model performance metric, we find that more knowledge transfer in terms of an increased number of transferred layers does not necessarily mean higher accuracy.}
    \SubItem {We assess the impact of the amount of target trace on the accuracy of the fine-tuned model, giving recommendations to this point.}

    \item{\textit{ContextPerf} automation tool}
    \SubItem {We automate the process of selecting a pre-trained source model, fine-tuning the model using the target
    environment trace, and generating synthetic traces using the fine-tuned model by prototyping a tool called ~\textit{ContextPerf}.}

    \item{Real mobile application QoE testing}
    \SubItem{We carry out example testing case-studies to estimate QoE metrics of \textit{Instagram} and \textit{Conversations} chat
    messenger mobile applications. The case-studies compare the QoE metric distributions obtained by using our synthetic access time
    traces with those obtained using the popular and simple normally distributed access time model.}

\end{itemize}
The paper is organized as follows: Cloud access time impact on various categories of mobile application QoE is discussed in  \sectionref{sec:qoe}.
\sectionref{sec:trace_generation} introduces the problem of modeling cloud access times using LSTMs and transfer learning.
\sectionref{sec:meth} and \sectionref{sec:eval} present the details of our methodology and performance analysis of transfer
learning to build cloud access time models. The synthetic traces generated using the
model are integrated with the mobile network emulator in  \sectionref{sec:emulation}.  \sectionref{sec:casestudies} describes mobile application
testing case-studies to measure QoE of two popular mobile applications using synthetic traces. The related work is discussed in  \sectionref{sec:related_work}.
Discussion of future work and challenges is presented in  \sectionref{sec:discussion}. Finally, we conclude in  \sectionref{sec:conclusions}.

\section{Cloud Access Times and the Mobile Application QoE}
\label{sec:qoe}

In this section, we discuss the quantified impact of cloud access times on the QoE for various categories of
mobile applications. Access time is measured as the time duration starting from when the first packet is sent
by the application task to the time when it reaches the destination cloud backend and vice versa
from the cloud backend to the application task. The interaction of an application task with the
cloud backend spans over multiple flows. Therefore, response times become additive and can
accumulate. A slight increase in single access flow latency can significantly impact the response time of
the application, thus adversely impacting the overall application QoE.

The severity of the impact depends on the application category and the action tasks within them.
For mobile web-based applications, Page Load Time (PLT) is one of the most important QoE metrics.
The changes in cloud access times impact PLTs. According to the study
in~\cite{180330}, a 25\% reduction in the
access time reduces the PLT by 45\%. Another study by Belshe~\cite{belshe}
shows that every 20~ms decrease
in access time leads to a linear decrease in the PLT. When the access times are higher than 100~ms, bandwidth increase
beyond 3 Mbps has almost no positive impact on the
PLT~\cite{10.1145/2872427.2883014}. Users can easily perceive the lags in web
page load
when access time suddenly increases by 100--200~ms. When the access time is above 300~ms, the page loads sluggishly,
and when it goes beyond 1~s, users move on (Grigorik
\etal~\cite{grigorik2013high}). Similar to PLT,
user perceived application latency is an important QoE metric for social media based mobile applications like Facebook.
The degradation in user perceived latencies can lead to frustrating experiences for its users.
Chen \etal~\cite{10.1145/2663716.2663726} find that access time is on the
critical path for some user actions like
photo sharing, and can contribute up to 65\% of the user-perceived latency. Another important measure of QoE is the user
application retention. A large-scale study by~\cite{10.1145/3308558.3313428} on
quantifying impacts of access times on
application retention rate found that the application retention rate can be sensitive to increase to access times across
all categories of mobile applications. They found that user retention rate (on a scale of 0--100) dropped from 98.3 to
66.5 when median access time increased from 34~ms to 79~ms for messenger based applications like Whatsapp and from 92.7 to
67.8 for a median access time increase from 37~ms to 70~ms for Twitter. Also, there have been significant efforts on the application
level protocols to reduce access time latency and improve the QoE. The newly
proposed HTTP/3~\cite{http3} standard uses the
QUIC transport protocol for the web. QUIC runs in the application layer on top of UDP, instead of TCP, therefore no additional
handshakes and slow starts are required, thus reducing the connection overhead. QUIC also introduces the concept of streams that
are delivered independently such that in most cases packet loss affecting one stream does not affect others. The literature
studies~\cite{7510788,7997281,10.1145/3131365.3131368} show that the PLTs using
QUIC are roughly 25\%--30\% faster than
HTTP/2 and 35\%--40\% faster than HTTP/1.1 in environments that have high access time latencies.

Besides the cloud access time distributions, the model should also be able to capture their time-based variations.
A study by Nikravesh \etal~\cite{DBLP:conf/pam/NikraveshCKMW14} shows that
significant non-uniform variations in
access times are possible across Mobile Network Operators (MNOs), geographic locations and different times of
day, thus making the access times difficult to model and predict. Certain mobile applications like Voice over IP (VoIP),
and online gaming are very sensitive to the large variation in cloud access times (jitter). A study of Skype users by
Chen \etal~\cite{10.1145/1159913.1159959} showed that the duration of VoIP
sessions is directly impacted by access
time and its jitter. The median duration of sessions with access times greater than 270~ms were 4~min, while sessions
with access times between 80~ms and 270~ms were 5.2~min. However, when access times dropped below 80~ms the session durations doubled to 11~min. In their empirical
model of user dissatisfaction, the access time jitter parameter has a weight factor of 53\%. A study by
Cisco~\cite{cisco_jitter} recommends 30~ms as the acceptable jitter threshold
for VoIP applications.

For online gaming applications, Wang \etal~\cite{5425784} define Game Mean
Opinion Score (GMOS) metric
for measuring their QoE. The GMOS metric has a range from 1.0 to 5.0, where 4.5--5.0 means an excellent game with no impairments, 3.0--4.0
implies a noticeable impairment, where the user might quit the game, and 1.0--2.0 means an annoying environment
where the user will definitely quit the game. They found that the GMOS in general decreases with an increase
in access time. Their measurement-based study of gaming experience on-campus WiFi network concludes that
although generally, GMOS greater than 4.0 is achievable on WiFi, there are also frequent periods of network instability
when access time suddenly can increase by ~200~ms which leads to a drop in user experience (GMOS-2.0). Pantel \etal
~\cite{10.1145/507670.507674} show that backend access time latencies over
100~ms can create paradoxical situations in
real-time multiplayer racing games. Studies in~\cite{10.1145/1167838.1167860}
for First Person Shooting (FPS) games
found that for backend access times above 100~ms, the hit rate of precision shooting reduced by more than 50\%, thus
putting certain players at disadvantage. In contrast, Real Time Strategy (RTS) games are more tolerant to backend
access times. Sheldonet al.~\cite{10.1145/963900.963901} study the effect of
backend access time on players of the
RTS game Warcraft III and conclude that while the latency of several seconds are noticeable for players, it has little
to no effect on the game outcome.

The above studies of various categories of mobile applications demonstrate the dependence and the sensitivity
of QoE on the access time distribution and their time-based variations and the need to model them accurately.

\section{Cloud Access Time Modeling}
\label{sec:trace_generation}
In this section, we discuss the background of our work by presenting the Long Short Term Memory (LSTM) neural net framework
which we have chosen for modeling cloud access times. Next, we introduce the transfer learning process used to build target environment
specific cloud access time models from pre-trained LSTM neural net models.

\subsection{The Long Short Term Memory (LSTM) Neural Net Framework}

Let a random variable $x_{t}$ represent the time required to access a real cloud service in a given
domain $\mathcal{D}$ at time $t$.
The domain here implies a combination of the end-user context,
a real mobile device, a real access network, and a real cloud backend service.
Given some constant measurement frequency, the real cloud
service access time traces for the domain $\mathcal{D}$ can be represented as a
discrete-time series vector $X_{D}=[x_{D1}, x_{D2}, \hdots ,x_{DZ}]$
of length $Z$ measured over a certain time period. A basic synthetic trace
generation problem can then be expressed as modeling a prediction function
$F$ that uses $X_{D}$ to generate the cloud access time $\hat{x}_{D(Z + 1 )}$ of
the domain $\mathcal{D}$ in the next time step $Z+1$.
\begin{equation}
    \hat y = \hat{x}_{D(Z + 1 )} = F(x_{D1}, x_{D2}, \hdots ,x_{DZ})
\end{equation}

The analytical framework that we apply to the trace generation problem is ``Long
Short Term Memory'' (LSTM) neural
net~\cite{Hochreiter:1997:LSM:1246443.1246450}.
LSTM is a state of the art deep learning model which can capture the temporal structures within the traces at different
resolutions, including both long-term as well as short-term. LSTM has been consistently used in the literature
for time series prediction
problems~\cite{DBLP:journals/corr/Graves13,Pascanu:2013:DTR:3042817.3043083,Sutskever:2014:SSL:2969033.2969173}
and proven to be stable and powerful.
\begin{figure}
    \includegraphics{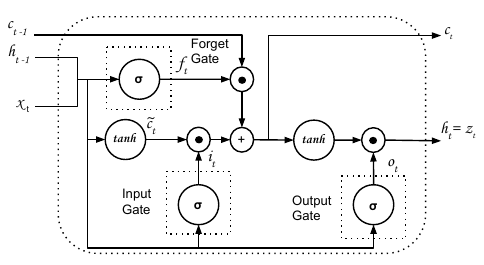}
    \caption{Typical Long Short Term Memory (LSTM) cell unit.\label{fig:lstm-architecture}}
\end{figure}

We use a typical LSTM cell unit shown in  \figureref{fig:lstm-architecture} and described by
the equations in~\ref{eq:lstm1}--\ref{eq:lstm}. $\sigma(\cdot)$ is a sigmoid function
that limits a real valued input between $\SIrange{0}{1}{}$ and $tanh(\cdot)$ represents the hyperbolic tangent function
that limits the real valued input in the $\SIrange{-1}{1}{}$ range. The $\odot$ represents element-wise
multiplication, and $t$ is the time step. The LSTM cell consists of a hidden state $h_{t} \in \mathbb{R}^N$ with N hidden units, an input
modulation gate $\widetilde{c_{t}} \in \mathbb{R}^N$, and three gates, the input gate $i_{t} \in \mathbb{R}^N$, the forget gate $f_{t} \in \mathbb{R}^N$, and the output gate $o_{t} \in \mathbb{R}^N$. The memory cell unit
$c_{t} \in \mathbb{R}^N$ accumulates the state information at time step $t$ and is the sum of the previous memory cell unit $c_{t-1}$ modulated by
$f_{t}$ and the input gate $i_{t}$ modulated by $\widetilde{c_{t}}$ which is a function of the current input $x_{t}$ and
previous hidden state $h_{t-1}$. Gates are a mechanism to optionally pass the input through them in order to update
the cell state. Whenever a new input arrives, its information is accumulated in the cell if the input gate $i_{t}$ is
activated. The prior cell status $c_{t-1}$ at step $t-1$ is forgotten if the forget gate $f_{t}$ is activated.
Similarly, the output gate $o_{t}$ learns how much of the memory cell to transfer to the hidden state $h_{t}$.
$W_{xj}$, $W_{hj}$, and $b_{j}$ where (j $=$ i, f, o, c) are the parameters of the cell unit that are learned
during the training.
\begin{align}
    i_{t} = \sigma(W_{xi}x_{t}+W_{hi}h_{t-1} + b_{i}) \label{eq:lstm1}\\
    f_{t} = \sigma(W_{xf}x_{t}+W_{hf}h_{t-1}+b_{f}) \\
    o_{t} = \sigma(W_{xo}x_{t}+W_{ho}h_{t-1}+b_{o}) \\
    \widetilde{c_{t}} = tanh(W_{xc}x_{t}+W_{hc}h_{t-1}+b_{c}) \\
    c_{t} = f_{t} \odot c_{t-1} + i_{t} \odot \widetilde{c_{t}} \\
    h_{t} = o_{t} \odot tanh(c_{t})
    \label{eq:lstm}
\end{align}

The network can make use of multiple LSTM units stacked on top of each other to make the network deeper. The stacking of
memory units on top of each other causes the hidden states to be propagated to the deeper layers, thus enabling the hierarchical processing of time series
data. An end-to-end network path can exhibit temporal dynamics at different time scales within the trace time series. Thus, stacking the layers
on top of each other may enable the processing of this temporal hierarchy.

\subsection{Transfer Learning}

\begin{figure*}
    \includegraphics{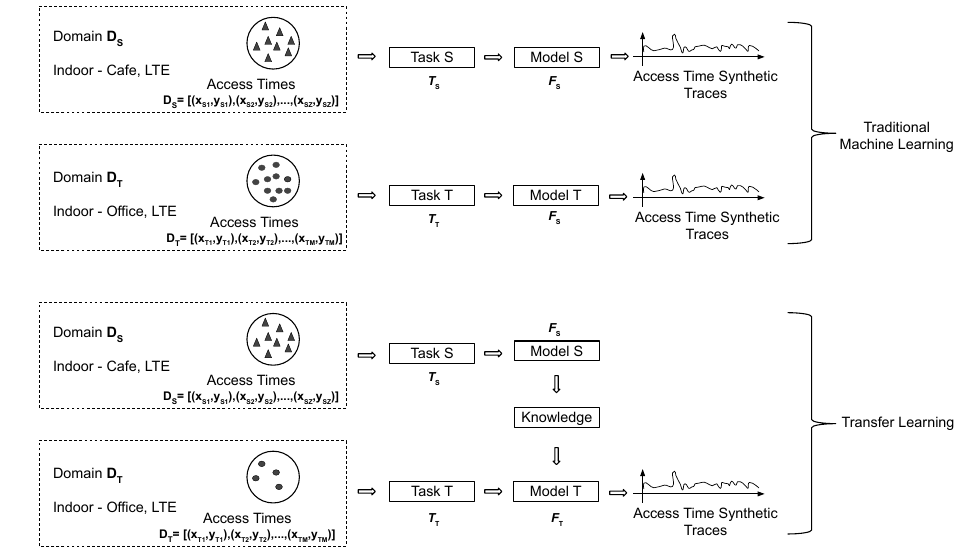}
    \caption{Traditional machine learning vs. Transfer learning approach to access time prediction.\label{fig:transferlearning}}

\end{figure*}

A domain $\mathcal{D}$ consists of an input feature space $\mathcal{X}_{D}$ and a marginal probability distribution $P(X_{D})$,
where $X_{D} =  [x_{D1}, x_{D2}, \hdots ,x_{DZ}]$ and $X_{D}\in\mathcal{X}_{D}$.
Thus, given a domain $\mathcal{D}=\{\mathcal{X}_{D},P(X_{D})\}$, and the output feature space $\mathcal{Y}_{D}$, the goal of a
task $\mathcal{T}_{D} = \{\mathcal{Y}_{D}, F_{D}(\cdot)\}$ is to learn a prediction function $F_{D}$ using the training data pairs
$D=[(x_{D1},y_{D1}), (x_{D2},y_{D2}), \hdots ,(x_{DZ},y_{DZ})]$, where $x_{Di}\in\mathcal{X}_{D}$ and $y_{Di}\in\mathcal{Y}_{D}$.
In the case of time series prediction problems, both the input and the output space are the same i.e $\mathcal{X}_{D} = \mathcal{Y}_{D}$.
Therefore, $x_{Di},y_{Di} \in \mathcal{X}_{D}$.
Further, since the prediction is carried out single step at a time, $y$ is obtained by shifting $x$ by one, i.e. $y_{i} = x_{i+1}$.
Thus, given an input access time $x$ at the current time step, the prediction function $F_{D}$ is
used to predict the output access time at the next time step. $F_{D}$ can also be expressed as a conditional probability
distribution function $P(y \vert x)$.

\figureref{fig:transferlearning} shows the comparison of the transfer learning approach with
the traditional machine learning approach for the access time prediction problem. We consider two domains: the source domain $\mathcal{D}_{S}$ (an indoor environment
such as a cafe with a stationary mobile device accessing a cloud backend service on a LTE network) and the
target domain $\mathcal{D}_{T}$ (an other indoor environment like an office with a different stationary mobile device accessing an other cloud backend over an other LTE network). In the source domain $\mathcal{D}_{S}=\{\mathcal{X}_{S},P(X_{S})\}$,
using the measured access times data pairs $D_{S}=[(x_{S1},y_{S1}), (x_{S2},y_{S2}), \hdots ,(x_{SZ},y_{SZ})]$ a source task $\mathcal{T}_{S}$ has
learned a prediction function $F_{S}$. We call this a pre-trained model (Model S in  \figureref{fig:transferlearning}).
Now in the target domain, $\mathcal{D}_{T}=\{\mathcal{X}_{T},P(X_{T})\}$ a short sample of the access time is measured.
This access time sample is represented by the vector $X_{T}=[x_{T1}, x_{T2}, \hdots ,x_{TM}]$ of length $M$ such that $0 < M \ll Z $.
Given the training data pairs $D_{T}=[(x_{T1},y_{T1}), \hdots ,(x_{TM},y_{TM})]$ such that $x_{Ti},y_{Ti} \in \mathcal{X}_{T}$,
transfer learning for access time prediction in $\mathcal{D}_{T}$ is defined by a target task $\mathcal{T}_{T} = \{\mathcal{Y}_{T}, F_{T}(\cdot)\}$.
In transfer learning, $\mathcal{T}_{T}$ learns a prediction function $F_{T}$ (Model T in  \figureref{fig:transferlearning}) using both
the training data pairs $D_{T}$ and also the knowledge in $D_{S}$ and $\mathcal{T}_{S}$ (pre-trained Model S). Further, we consider
$D_{S}\neq D_{T}$, since the marginal probability distributions of the source and the target domains are different i.e $P(X_{S})\neq P(X_{T})$.

\section{Synthetic Access Time Trace Generation}
\label{sec:meth}
In this section, we present the methodology for
using transfer learning to address the problem of modeling cloud access
times for target environments. First, we explain the details of context-driven network testing scenarios and cloud access time data collection, and our archive of cloud access time traces. Next, we explain our approach to building the source model library.
The pre-trained model library forms the basis for our transfer learning to enable
rapid model building and trace generation of target environments.
Based on the earlier transfer learning studies in other
domains~\cite{Weiss2016}, we hypothesize that transfer
learning performance is highly impacted by the choice of the pre-trained
source model used for learning, the number of layers to be transferred in the learning, and the amount of
target environment data available for learning. Therefore, we carefully design experiments to characterize
transfer-learning performance of cloud access time models under the impact of these factors.

\subsection{Cloud Access Time Trace Collection}
\label{sec:tool}

Since there are no readily available pre-trained LSTM models for mobile cloud access times, we build them ourselves.
We use the LSTM neural net architecture framework consisting of a stack of LSTM cell units as our pre-trained source models.
Furthermore, there are no publicly available datasets for mobile cloud access time series. The access times and the
network performance has been known to exhibit a high degree of correlation to the end-users
location, environment, and the situation~\cite{DBLP:conf/imc/GemberAPVC12}; the
factor we refer to as context in this paper.
Thus, these access time measurements need to be labeled with appropriate context information that could then
be used to build LSTM models. Inspired by the previous work
Imagenet~\cite{DBLP:journals/cacm/KrizhevskySH17} in the machine
learning domain, we build an archive of cloud access time traces for
a selected set of contexts.

\begin{figure}
    \centering
    \includegraphics[width=\columnwidth]{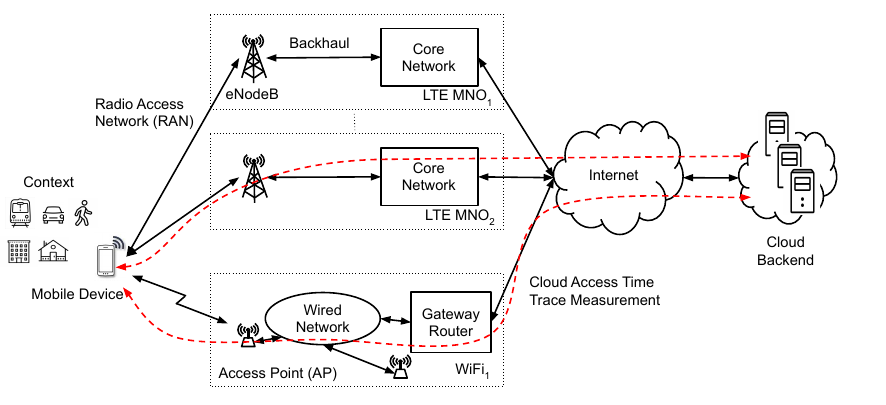}
    \caption{Experimental setup for collecting cloud access time traces on the LTE and WiFi networks for different contexts. The end-to-end
    network path for cloud access time measurement is shown with the red dashed line.
    \label{fig:measurement-setup}}
\end{figure}

\begin{table*}
    \centering
    \caption{Overview of the cloud access time dataset archive showing different contexts, locations (DE - Germany, US - United States), network types (LTE, WiFi) and mobile network operator (MNO) on which measurements were carried
    and their respective RTT counts.}
    \begin{tabular}{|l|l|l|l|l|l|}
        \hline
        \multicolumn{2}{|c|}{\textbf{Context}} & \multicolumn{4}{c|}{\textbf{Cloud access time dataset}}\\\hline
        \multirow{2}{*}{\textbf{Type}} & \multirow{2}{*}{\textbf{Scenario}}        & \multicolumn{2}{c|}{\textbf{LTE}} & \multicolumn{2}{c|}{\textbf{WiFi}} \\\cline{3-6}
        &                                  & Location       & Count   & Location & Count \\\hline
        \multirow{4}{*}{Mobile}      & Train                     & DE ($MNO_{1}$)   & 30K     &  \multirow{2}{*}{-}         & \multirow{2}{*}{-} \\\cline{2-4}
        & Vehicle                   & DE ($MNO_{1}$)   & 30K     &          &  \\\cline{2-6}
        & \multirow{2}{*}{Walking}  & US ($MNO_{2}$)        & 30K     &   \multirow{2}{*}{DE (Campus)}       &  \multirow{2}{*}{30K} \\\cline{3-4}
        &                           & DE ($MNO_{3}$)   & 30K     &          & \\\hline
        \multirow{5}{*}{Stationary}  & \multirow{4}{*}{Indoor}   & \multirow{4}{*}{DE ($MNO_{1}$)}   & \multirow{4}{*}{30K}     &  DE (Home)        & 30K \\\cline{5-6}
        &                           &                             &                         &  DE (Office)      & 30K \\\cline{5-6}
        &                            &                           &                             &  DE (Dorm)              & 30K \\\cline{5-6}
        &                            &                             &                         &  US (Cafe)      & 30K \\\cline{2-6}
        & Outdoor                   & DE ($MNO_{3}$)   & 30K     &     DE (Cafe)     & 30K \\
        \hline
    \end{tabular}
    \label{tab:domain}
\end{table*}

\begin{figure*}
    \centering
    \includegraphics[width=0.8\textwidth]{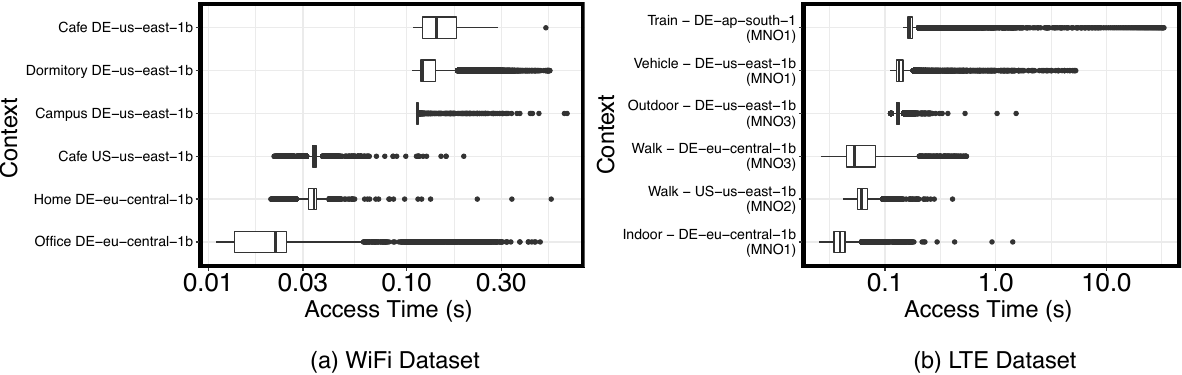}
    \caption{Boxplots of our WiFi and the LTE datasets consisting of 30\,000 cloud access time (RTT) samples for different contexts. The box in the figure is defined by the 25th, 50th, 75th percentiles.
    And the access times are plotted on base 10 logarithmic scale.\label{fig:datasets}}
\end{figure*}

\subsubsection{Measurement Setup and Scenario}

Our experimental setup for collecting cloud access time measurements is shown in  \figureref{fig:measurement-setup}.
We  collect cloud access time measurements for a combination of a mobile device, cloud backends, and access networks for
various end-user contexts.
The different contexts include the mobile scenarios of users in the outdoor urban environment with walking,
vehicle, and train commutes. We also considered stationary end-user scenarios at various urban locations in
indoor environments such as home, office, dormitory, and cafe.
It is common for mobile applications to use backend servers hosted in the public cloud. Therefore, we use AWS to host the
backend, as it is well provisioned with high availability across different geographical regions and it is easier to isolate
and model the impact of access times. The application backend server is hosted on a Virtual Machine (VM) in three different
geographical regions of AWS data-center --- United States (Virginia), Europe (Frankfurt), and Asia Pacific (Mumbai).
The access network connects the mobile device to the cloud backend. For our measurements, we consider WiFi networks and LTE cellular networks from different Mobile Network Operators (MNOs).

The scope of our data collection is in no way extensive. The context environments were selected based on common mobile
application network testing scenarios. On the other end, the access networks are limited to LTE and WiFi.
These data collection could be extended as needed for a given class of access networks, end-user
locations and contexts, and the cloud backends geographically distributed over various regions.
In fact, we plan to expand their scope to include 5G as a part of our future work.

For each cloud access time measurement, we fix the context, connect the device to a selected network, and select an application backend.
In the case of WiFi, we have collected measurements in total on six different networks including public and private WiFi networks.
In order to collect measurements on cellular networks, we have used LTE from three different Mobile Network Operators (MNOs) in two
countries --- United States (US) and Germany (DE).

\subsubsection{Cloud Access Time Measurement}

We define the cloud access times as the network Round Trip Time (RTT) between the mobile device and the cloud backend.
The network RTTs are measured for each context mentioned in~\tabref{tab:domain}. The RTTs are measured
using our prototyped \textit{ContextPing} Android application. \textit{ContextPing} uses the ported standard native UNIX ping utility tool
based on ICMP. It enables the configuration of the cloud application backend to which the access time should be measured, the
frequency with which ICMP packets are sent, and the total time duration of the measurement. Each of our RTT time series measurement dataset
lasts for a time duration of 15\,000~s (4~h and 10~min) and consists of 30\,000 ICMP packets generated
with a time interval of 500~ms. All our cloud access time datasets were collected using a Samsung A70 mobile
phone that ran \textit{ContextPing} on the Android 9.0 Operating System. An overview of our trace collection is shown in \tabref{tab:domain}. The boxplots for the WiFi and the LTE cloud access time datasets are shown in  \figureref{fig:datasets}a and~b, respectively. In the remainder
of the paper, we refer to each of these 30\,000 RTT measurement samples as a single dataset.

\subsection{Building Source Model Library}

\subsubsection{Performance Metric}

There are several performance metrics such as Root Mean Squared Error (RMSE), Mean Absolute Error (MAE), Mean Absolute Percentage
Error (MAPE), Symmetric Mean Absolute Percentage Error (SMAPE), etc. that are used for evaluating the accuracy of trace models that
generate time series data (see~\cite{Chen04assessingforecast}). Both RMSE and
MAE are unit dependent measures. MAPE is the most
widespread metric for evaluating time series prediction accuracy. However, it places a heavier penalty on predicted values that
exceed the real value~\cite{ARMSTRONG199269}. Further, it biases the estimate
if the real values are small or zero.
Symmetric Mean Absolute Percentage Error (SMAPE) introduced by Makridakis
\etal~\cite{MAKRIDAKIS1993527}
is more balanced in handling smaller real values. Although the name suggests otherwise, SMAPE is asymmetric by nature while MAPE is
symmetric (see~\cite{GOODWIN1999405}). SMAPE has been widely used in
time series prediction problems like
demand prediction during extreme events at Uber~\cite{uber}, predicting
electricity load during peaks~\cite{transferlearning} etc.
Therefore, we use a scaled version of the SMAPE performance metric proposed
in~\cite{Chen04assessingforecast} for our pre-trained source
model evaluation. It is defined as  follows:
\begin{equation}
    SMAPE=\frac{100\%}{N}\sum_{i=1}^{N}\frac{\abs{y^{\prime}_{i}-y_{i}}}{\abs{y_{i}}+\abs{y^{\prime}_{i}}}
\end{equation}
where $y^{\prime}_{i}$ is the model predicted value of the $i_{th}$ data-point in the time series and $y_{i}$ is
the real value.

\subsubsection{Building Source Models}
\label{sec:source_models}

\begin{figure}
    \centering
    \includegraphics[width=\columnwidth]{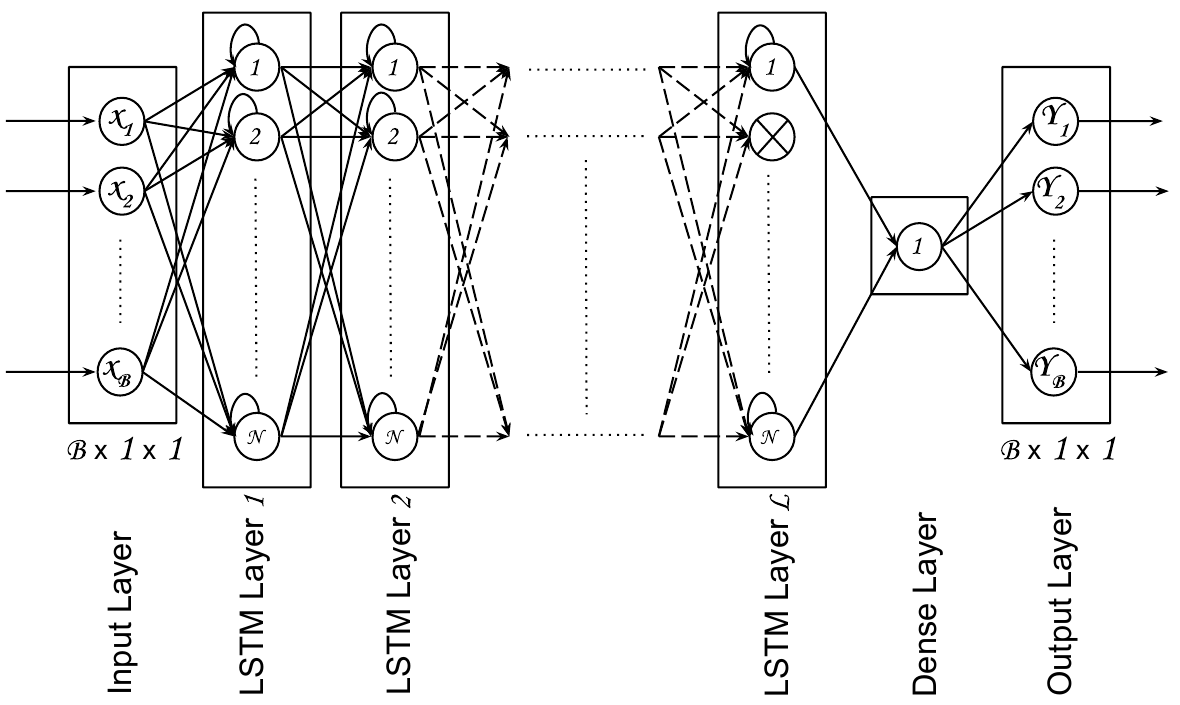}
    \caption{Our LSTM based source model architecture.\label{fig:lstm-architecture-stack}}
\end{figure}

\textbf{Source Model Architecture --- }
Our goal is to find a specific instance of the source model architecture for each of the selected contexts. We use the general
architecture shown in  \figureref{fig:lstm-architecture-stack}. The input layer is connected to \textit{L} stacked layers of
LSTM cell units followed by the dense layer and the output layer. The number of stacked layers of LSTM cell units
(\textit{L}) can vary for each source dataset. \textit{N} is
the number of hidden units in each of the LSTM units. Overfitting can be a serious problem in training
LSTM based architectures, therefore a dropout function is commonly used to avoid overfitting. A dropout function
randomly drops hidden units during the training, and thus allows different units to learn and prevents co-adaptation
among them. It is known to improve neural net
performance~\cite{journals/jmlr/SrivastavaHKSS14}. Therefore, in the
final LSTM layer \textit{L}, a dropout function is applied. A dense layer is an activation function that maps the multi-dimensional input received after the dropout in the last LSTM layer to the output which is
the predicted value of access time. An activation function should be able to output values that are
far from the mean and median values with higher probabilities, and thus preserve the long tail nature of access time RTTs.
An activation function like Exponential Linear Unit
(ELU)~\cite{clevert2016fast} satisfies this requirement. The ELU function
is defined as $x$ if $x > 0$ and $\alpha \cdot (exp(x) - 1)$ if $x < 0$. Thus, ELU is an identity function for positive inputs. And for the
negative inputs, the output is smoothed towards the hyperparameter $\alpha$. Thus, $\alpha$ controls
the value to which an ELU saturates for negative inputs. However, ELU is deterministic by nature and exhibits a fixed
input-output relationship, thus limiting its generalization ability. In order to obtain better generalization, we add
a stochastic perturbation $\sigma\epsilon$ to ELU, where $\sigma$ is the range of stochastic perturbation and $\epsilon$
is randomly sampled from \textit{$\mathcal{N}$(0,1)}. This stochastic activation
function was proposed by Lee \etal~\cite{lee2019probact}
and is called Probabilistic Activation (ProbAct). In our ProbAct activation function, an ELU function is used with $\alpha$ $=$ 1.0,
$\sigma$ $=$ 1.0. The input layer takes as input a single data-point $x_{i}$ of the cloud access time series and the output layer
outputs the predicted value $y_{i}$ which is the next time step in the time series. For training, the input layer is a
tensor of dimension \textit{B} \time1\time1, where \textit{B} is the batch size parameter that would be used in training
this architecture and is explained later.

\textbf{Training and Testing Data Splits --- }
We split each of the cloud time series datasets,
i.e.,~$[x_{1}, \ldots .,x_{Z}]$ into two subsets preserving the ordering of the time series. The training set, $S_{Training}$,
consists of $x_{1}, \ldots .x_{P}$ (where $P<Z$) and is used for training the architecture. The testing set,
$S_{Testing}$, consists of $[x_{P+1}, \ldots .,x_{Z}]$  and is used for testing. This split is made in an 80:20 ratio
such that 80\% samples (24\,000 RTT pings) in the time series are in $S_{Training}$ set and the remaining 20\% (6000 RTT pings)
in the $S_{Testing}$ set.

\textbf{Preprocessing --- }Data cleaning and processing is a precursor step to training a model. In this step,
we standardize the $S_{Training}$ and $S_{Testing}$ datasets separately. Standardization is defined as subtracting
the mean value and dividing by the standard deviation of the dataset from each data point in the
dataset. Standardization makes the dataset scale-invariant, thus enabling the model to generate cloud
access time traces across multiple data scales.

Next, we transform the $S_{Training}$ and $S_{Testing}$ datasets into labeled datasets such that it is suitable
for training LSTM architecture in a supervised manner. This means for every data-point (input feature)
in the input $S_{Training}$ time series, we have an output data-point (output feature).
Since our LSTM architecture predicts one step at a time, the output ground truth time series is obtained by
shifting the input time series by a factor of one. Each data-point $y_{i}$ in the output time series is the $x_{i+1}$
data-point in the input time series. Thus, the $S_{Training}$ and $S_{Testing}$ datasets are transformed into
tensors of $P \times 2 \times 1$ and $(Z-P) \times 2 \times 1$
dimensions, respectively.

\textbf{Training --- }
The goal of the training is to identify suitable weights that minimize a loss function
using the transformed labeled $S_{Training}$ dataset. We use the Mean Squared Error (MSE) between the estimated
and the actual data-points in the output time series of the labeled $S_{Training}$ dataset as the loss function
for our training.

For training, we use the gradient descent based optimization algorithm based on adaptive learning
rates called Adam proposed by Kingma \etal~\cite{kingma2014adam}. The Adam
optimizer is described in
Algorithm 1 and can be explained as follows. $f(\theta)$ is the stochastic objective function whose expected value $E[f(\theta)]$ is
to be minimized with respect to parameters $\theta$. $f_{1}(\theta) \cdot\ldots\cdot f_{T}(\theta)$ denote the value
of the function over each time-step $1, 2,\ldots,T$. And $g_{t} = \nabla_{\theta}f_{t}(\theta)$ denotes the gradient, i.e.~the vector of partial derivatives of $f_{t}$ with respect to $\theta$ at time-step $t$. The algorithm evaluates the exponential
moving averages of the gradient ($m_{t}$) and the squared gradient ($v_{t}$) at  each time step $t$. The hyper-parameters
$\beta_{1}, \beta_{2} \in [0,1)$ control the rate of exponential decay of moving averages. These moving averages are the
estimated 1st moment (the mean) and the 2nd raw moment (the uncentered
variance) of the gradient. Since the moving averages are initialized to zeros, their estimates are biased towards zero. This is particularly noticeable
during the initial time-steps and when the decay rates are small. Therefore, a bias correction factor is applied to obtain
$\hat{m}_{t}, \hat{v}_{t}$. The bias corrected estimates are then used to take a time-step $\Delta_{t}$ in parameter space
which is equal to $\alpha \cdot \hat{m}_{t}/(\sqrt{\hat{v}_{t}}+\epsilon)$. As the learning rate $\alpha$ is responsible for selecting
the magnitude of these parameter space steps, it can be adapted in such a way so that the optima is reached in few iterations.

\begin{algorithm}
    \SetAlgoLined

    \DontPrintSemicolon
    \kwInput{$f(\theta)$ with parameters $\theta$}
    \Numberline \kwInit{$\theta_{0}, m_{0}=0, v_{0}=0, t=0$}
    \While{$\theta_{t}$ not converged}{
    $t \leftarrow t + 1$\;
    $g_{t} = \nabla_{\theta}f_{t}(\theta_{t-1})$\;
    $m_{t} \leftarrow \beta_{1}m_{t-1} + (1 - \beta_{1})g_{t}$\;
    $v_{t} \leftarrow \beta_{2}v_{t-1} + (1 - \beta_{2})g_{t}^{2}$\;
    $\hat{m}_{t} \leftarrow m_{t} \mathbin{/} (1 - \beta^{t}_{1})$\;
    $\hat{v}_{t} \leftarrow v_{t} \mathbin{/} (1 - \beta^{t}_{2})$\;
    $\theta_{t} \leftarrow \theta_{t-1} - \alpha \cdot \hat{m}_{t}/(\sqrt{\hat{v}_{t}}+\epsilon)$\;
    }
    \KwRet{$\theta_{t}$}
    \caption{Adam Optimization as proposed in~\cite{kingma2014adam}. Symbol $g_{t}^{2}$ indicates the elementwise square $g_{t} \odot g_{t}$. Learning rate is denoted as $\alpha$
    and set to 0.00001. Hyper-Parameters are set to $ \beta_{1} = 0.9$, $ \beta_{2} = 0.999$, and  $\epsilon = 10^{- 7}$}
\end{algorithm}

Further, we train in batches of size \textit{B} since the entire $S_{Training}$ data may not fit in the memory in a single run.
We use different batch sizes (B) in training from 4 to 32 in increments of power of 2.
The batched data is fed without shuffling to the LSTM architecture, thus preserving the order of data-points.
We also ensure that each LSTM cell retains its state while training across the batches as we have a
single time series input that spans multiple batches. In order to get the optimizer to converge, we train the
architecture over multiple cycles through the entire training dataset. Each such cycle is referred to as the epoch.
We vary the number of epochs from 100 to 700 in increments of 100.

\begin{table}[htp]
    \centering
    \caption{Hyperparameters of the architecture and training}
    \centering
    \begin{tabular}{|c|c|c|}
        \hline
        \textbf{Type} & \textbf{Hyperparameter}  & \textbf{Space} \\\hline
        \multirow{7}{*}{Architecture}& Number of stacked & \multirow{3}{*}{1, 2, 3, 4 }\\

        & layers of LSTM  & \\
        & cell units (\textit{L}) & \\
        \cline{2-3}
        & Hidden units (\textit{N}) & 8, 16, 32, 64, 128, 256, 512\\\cline{2-3}
        & Dropout                    & 0.5\\\cline{2-3}
        & Dense layer & ProbAct~\cite{lee2019probact} with ELU $\alpha$=1.0\\
        &   activation function           & perturbation $\sigma$=1.0 $\epsilon$=\textit{$\mathcal{N}$(0,1)}\\\hline
        \multirow{3}{*}{Training} & Batch Size (\textit{B})    & 4, 8, 16, 32\\\cline{2-3}
        & Epochs                     & 100 - 700\\\cline{2-3}
        & Learning Rate              & $\alpha$=0.00001\\\hline
    \end{tabular}
    \label{tab:hyperparameters}
\end{table}

\textbf{Hyperparameters Tuning \& Testing --- }
The hyperparameters (highlighted in \tabref{tab:hyperparameters}) of the architecture
include the number of layers of stacked LSTM cell units (\textit{L}). We vary the number of stacked layers
of the LSTM units in the architecture from 1 to 4. The number of hidden units (N) in each LSTM cell unit of the stacked layers are varied
from 8, 16, 32, 64, 128, 256, 512. The dropout function value is set to 0.5
which has been shown by Srivastava \etal~\cite{journals/jmlr/SrivastavaHKSS14}
as being close to the optimal for a wide range of networks and training tasks. After carrying out the hyperparameter
tuning, we select the parameterization with the minimum MSE loss on the $S_{Testing}$ set as the pre-trained source model.
Thus, in the end, we have a separate source model for each dataset in the archive. Each of these source models can
have different numbers of stacked layers and also differ in the values of other hyperparameters.

\textbf{Implementation --- }The source model architecture is implemented in Python using the open-source deep-learning
library Keras~\cite{keras} and uses the Tensorflow backend. Our source model
training is executed in the AI Platform
service from the Google Cloud Platform (GCP). We run our experiments on a GCP predefined scale tier cluster
consisting of a \textit{n1-standard-4} instance type and includes a single worker instance. We found this tier to be
suitable for training the source models after initial experimentation. The total cumulative duration
for the model training, including the hyperparameter tuning, across all source datasets, takes around
three days.

\subsection{Transfer Learning for Target Environment Model Generation}

In the following, we describe the experiments designed to evaluate the robustness of
transfer learning for various targeted context environments on WiFi, and LTE networks.

\subsubsection{Selection of Pre-trained Models for Fine-tuning}

Earlier studies~\cite{Weiss2016} have shown that within a specific architecture,
using transfer learning with fine-tuning
of arbitrarily chosen source model can not only improve
but also worsen the model accuracy compared to when the model is trained from scratch. This phenomenon is known
as a positive and negative transfer, respectively. Therefore, we need to select the source models for fine-tuning
in an informed manner. The source model selection criteria should also make it easy for application testers to quickly
select the optimal source model for the target environment of interest. The
authors in~\cite{wang2018characterizing} show that
selecting a source model whose dataset characteristics are similar to the target dataset characteristics can increase the likelihood of positive transfer learning.
Therefore, we use a similarity distance measure for time series called Dynamic Time Warping (DTW)
proposed in~\cite{DBLP:journals/corr/abs-1811-01533} for selecting the source
model. DTW is an elastic
measure that optimally aligns the time series in the temporal domain. Thus, it also works when the two
time series have different lengths and have been sampled with different frequencies.

\subsubsection{Transfer Learning Organization}

The Target Domain ($D_{T}$) experiment consists of using transfer learning to fine-tune a model from pre-trained Source
Domain ($D_{S}$) with the access time sample of $D_{T}$. We split the $D_{T}$ trace sample into two sets,
$T_{Training}$ and $T_{Testing}$, which are respectively used for training and testing the $D_{T}$ model.
As earlier, we also standardize the target datasets separately. Next, we select a
pre-trained $D_{S}$ model and freeze the weights in the initial LSTM layers of the architecture (starting from LSTM
layer one) and then fine-tune on the remaining LSTM layers. We hypothesize that when stacking the LSTM layers to build
a $D_{T}$ model, the weights in the initial layers are more representative of the learning of the generic features
present in time series data. Thus, these features learned on the $D_{S}$ can be used in the $D_{T}$. On the other
hand, the final layers in the architecture are closely related to the predictive component of the model. We transfer
the weights of the initial layers in $D_{S}$ by freezing them and fine-tune the model using the target $T_{Training}$ set on the
final layers. Thus, we transfer the learnings carried out in the $D_{S}$ model into fine-tuning the $D_{T}$ model. The $D_{T}$ model is tested using
$T_{Testing}$. We study the impact of the following factors on the transfer learning performance.

\textbf{Impact of the source and target dataset similarity} --- To study the impact of source and target dataset similarity on
the transfer learning performance, we carry out transfer learning between each pair of the dataset in our archive.
We use the source models trained in the earlier step for transfer learning. The DTW distance measure characterizes
the dataset similarity.

\textbf{Impact of the number of transferred layers} --- In transfer learning to obtain a target model, fine-tuning is carried
out only on a subset of layers from the pre-trained source model while retaining the weights of the remaining subset of
layers by freezing them. We also study the impact of the number of transferred layers on the transfer learning performance.

\textbf{Impact of the amount of target data} --- To evaluate the impact of the amount of target environment data
available for fine-tuning the model, we vary the size of the target dataset used in fine-tuning.

\subsubsection{Performance Metric for Transfer Learning}

For a target environment, we introduce the notion of a specialized model. A specialized model is obtained by
training from scratch on the target environment data as if no pre-trained source model existed. A specialized model
can be obtained by using the training process defined in  \sectionref{sec:source_models}. The specialized models provide the baseline performance for transfer learning. As a performance metric, we consider the percentage improvement in the SMAPE of the fine-tuned model compared to a specialized model. It is computed as follows:
\begin{strip}
\begin{equation}
    \frac{\left( SMAPE_{specialized-model} - SMAPE_{fine-tuned-model} \right)}{SMAPE_{specialized-model}}\times 100
\end{equation}
\end{strip}

A positive value of SMAPE improvement percentage indicates that a fine-tuned model is better than training the
specialized model. On the other hand, a negative value indicates that the fine-tuned model is worse than the
specialized model.

\subsubsection{Analysis Methodology}

In the following, we briefly describe how we analyze the transfer learning experiment results.
To study the impact of source and target dataset similarity on transfer learning, we carry out
transfer learning across all pairs of datasets in our archive by transferring the initial
LSTM layers of the pre-trained source model architecture. This gives us a set of fine-tuned models. Then, for each
target environment, we also train a specialized model. Finally, we also compute the DTW for each dataset pair and
study if there exists a correlation between the DTW distance measure and the transfer learning performance using the
SMAPE improvement metric.

For studying the impact of the number of transferred layers on the transfer learning performance,
we select the pre-trained source model based on minimum DTW distance. The initial LSTM layers in the architecture
are known to represent the generic features of the access time series data that are not specific to a particular dataset
and in general applicable across different
datasets~\cite{10.5555/2969033.2969197}. Therefore, for this study,
we carry out transfer learning by varying the number of transferred layers from 1 to 3 beginning from the
first LSTM layer to the last LSTM layer in the architecture. For each transferred layer, we carry out ten runs of
out-of-sample validation typically used in time series modeling.

To study the impact of the size of target environment data used in fine-tuning on the transfer learning performance, we split
the target data into different ratios consisting of 20:80, 40:60, 60:40,
and 80:20 for training and testing, respectively. We then use the training subset to fine-tune the pre-trained source model.
The pre-trained source model for each target environment is selected based on the minimum DTW similarity measure. In each of the fine-tuning,
we set the number of transferred layers to one, and carry out ten runs of out-of-sample validation.

\section{Results and Analysis}
\label{sec:eval}

\begin{figure*}
    \centering
    \includegraphics[width=0.9\textwidth]{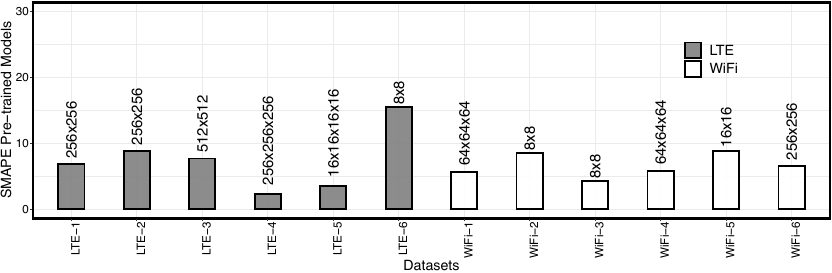}
    \caption{Bar plots illustrating SMAPE of the source model architecture evaluated on the test data of each WiFi and LTE dataset in the archive.\label{fig:barplot-pretrained-models}}
\end{figure*}

In the following, we present our results across the six WiFi and the six LTE datasets in the archive obtained using the
described methodology.

\subsection{Pre-trained Source Model}

In this section, we describe the results of training the source models.  \figureref{fig:barplot-pretrained-models} shows the results of each of the six
WiFi and LTE datasets, respectively. For each case, we report the SMAPE performance metric on the
20\% testing dataset. The diversity in the SMAPE performance metric within the WiFi and the LTE datasets is visible
in the figure. The SMAPE of LSTM architecture for the WiFi datasets are in the range of 5\% to 8.1\% while that for LTE datasets
vary between 2.3\% to 15.4\%. All source models except the one for LTE have SMAPE below 10\%, thus
showing a good prediction power for the model.

The text on the top of the bar indicates the source model architecture selected for each dataset. For example, 8~\ttimes~8~\ttimes~8 implies a stack of
three-layered LSTM architecture with 8 hidden units in each of the layers. The dropout layer and the dense layer in the architecture
are not included in the notation for the sake of simplicity.
We can see in  \figureref{fig:barplot-pretrained-models} that the best-fit source model
of every dataset has a different number of stacked LSTM layers. In the case of WiFi datasets,
the number of stacked LSTM layers vary from 2 to 3, while for the LTE datasets, the number of LSTM
layers vary from 2 to 4. Likewise, the number of hidden units in each of these LSTM architectures are also different.
We also analyzed each of the datasets and their LSTM architectures representing them and found that datasets with
longer tails have comparatively deeper LSTM architectures.  In our experiments, we observed that the MSE loss function of the
training stops improving after 700 epochs and that using a batch size of 16 in the training leads to minimal loss value.
We use these architectures as pre-trained source models in our transfer learning experiments.

\subsection{Transfer Learning}

To understand the underlying factors impacting the performance of transfer learning, we first focus on
characterizing the influence of pre-trained source model selection. Next, we study factors such as the number
of transferred layers and the size of the target data used in the model fine-tuning.

\subsubsection{Pre-trained Source Model Selection for Fine-tuning}

\begin{figure*}
    \centering
    \includegraphics{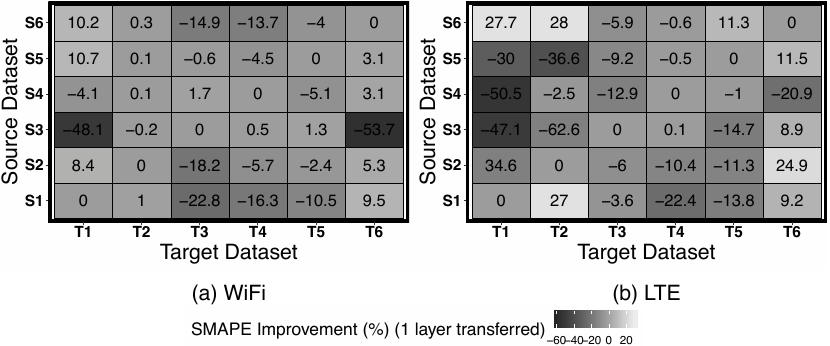}
    \caption{Positive and negative transfer learning between the source (S) and target (T) datasets. Heatmap illustrates
    percentage of SMAPE improvement in the target model (obtained with fine-tuning a source model) compared to
    training the specialized model from scratch.\label{fig:heatmap}}
\end{figure*}

\begin{figure*}
    \centering
    \includegraphics{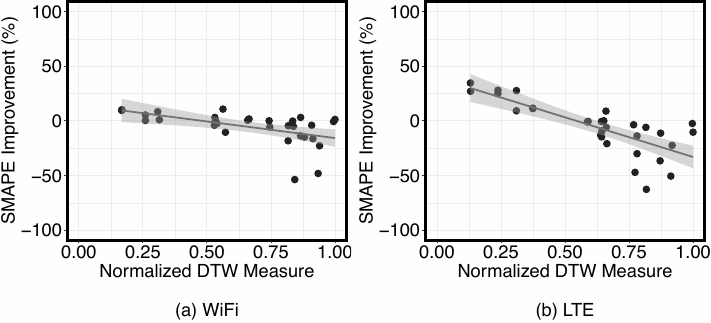}
    \caption{Scatter plot shows SMAPE improvement in the target model against the normalized DTW measure between the source (S)
    and target (T) datasets. The Pearson correlation coefficient between the normalized DTW measure and the SMAPE improvement
    for the WiFi dataset is \tmin0.46 and for the LTE dataset it is \tmin0.61,      thus showing moderate to strong negative correlation.\label{fig:dtw}}
\end{figure*}

We carry out transfer learning across each pair of the datasets in the archive (introduced in  \sectionref{sec:tool}) by fine-tuning
the source model while treating other datasets as the target data. Thus, we fine-tune 30 models for the WiFi and LTE each.
In the fine-tuning, the single first layer from the pre-trained source model is transferred, while the higher layers are fine-tuned.

\figureref{fig:heatmap}a and~b illustrate the heatmap of SMAPE improvement obtained from fine-tuning
a source model against just training the specialized model from scratch for WiFi and LTE datasets respectively.
The values along the diagonal from bottom left to top right can be ignored as they represent the same datasets.
The values inside the heatmap tile depict the percentage of SMAPE improvement in the target model obtained from fine-tuning the
source model. Positive values indicate that the SMAPE accuracy of the target model after fine-tuning using transfer learning
is better than that of the specialized model trained from scratch, while the negative values imply that the
fine-tuned target model is worse than the specialized model.

In the case of WiFi-based target models, we observe SMAPE improvements in only 46\% of the cases out of 30, while for
the LTE-based target models, this value drops to 33\%. The positive SMAPE improvements for the WiFi models
range from 0.1\% to 10.7\% and are marginal in the majority of the cases.
In comparison, the LTE models exhibit a much wider range of SMAPE improvement from 0.1\% to 34.6\%.
From the negative SMAPE improvement results, we can confirm that as expected using transfer learning is not always beneficial.
The positive SMAPE improvement results show that selecting certain source models can provide
marginal to high accuracy gains. However, randomly selecting a source model based only on the network
technology type, i.e.,~WiFi and LTE, may not offer any particular learning benefits. These results are in agreement with
Wang \etal~\cite{wang2018characterizing} and show that negative transfer
results can be caused by
a divergence in the joint distribution of the source and target datasets.

Based on the study in~\cite{DBLP:journals/corr/abs-1811-01533}, we use the
Dynamic Time Warping (DTW) distance measure for
establishing similarity between the underlying source and target datasets.  \figureref{fig:dtw}a (WiFi) and~b (LTE)
show the scatter plots of the SMAPE performance improvement for the transfer learning between each pair of the dataset and
the normalized DTW distance measure between their underlying source and target datasets. The scatter plots
show that SMAPE improvement percentage indeed decreases with a decrease in the similarity between the source and target datasets.
The Pearson correlation coefficient between the normalized DTW and the SMAPE improvement is \tmin0.46 and \tmin0.61 for the WiFi and the
LTE, respectively. The percentage of positive transfer cases jump from 46\% to 83\% for the WiFi and 33\% to 66\%
for the LTE when the minimum DTW measure is used to select the pre-trained source model. From the above results, we conclude
that a similarity measure like DTW is quite reliable, and using it can significantly improve the likelihood of
positive transfer.

\subsubsection{Transfer Learning Scope}

\begin{table}
    \centering
    \caption{Impact of the number of transferred layers used in fine-tuning a pre-trained source model on the target model performance.
    The Not Applicable (NA) indicates that the specific architecture is shallower than the number of transferred layers.}
    \centering
    \begin{tabular}{|c|c|c|c|c|c|}
        \hline
        \multirow{ 3}{*}{\textbf{Type}} & \multirow{ 3}{*}{\textbf{Target}} & \multirow{ 3}{*}{\textbf{Pre-trained Source}} & \multicolumn{3}{|c|}{\textbf{Average SMAPE}}\\
        & & & \multicolumn{3}{|c|}{\textbf{by number of}} \\
        &  \textbf{Dataset} & \textbf{Model} & \multicolumn{3}{|c|}{\textbf{transferred layers}} \\
        \cline{4-6}
        &                &                            & 1 & 2 & 3\\\hline

        \multirow{ 6}{*}{\textbf{WiFi}}&1&       $8\times8$                     & 5.5 & NA & NA\\
        &2&       $256\times256$                      & 5.1 & NA & NA\\
        &3&       $64\times64\times64$                     & 5.7    & 10.0 & NA\\
        &4&       $8\times8$                          & 5.8    & NA & NA\\
        &5&       $256\times256$                      & 6.1    & NA & NA\\
        &6&       $64\times64\times64$                     & 5.7    & 8.1 & NA\\\hline
        \multirow{ 6}{*}{\textbf{LTE}}     &1&       $256\times256$                     & 6.8  & NA & NA \\
        &2&       $256\times256$                         & 4.7  &  NA    &  NA \\
        &3&       $256\times256\times256$                     & 3.3 & 4.1 & NA\\
        &4&       $16\times16\times16\times16$                     & 2.7  & 2.5 & 4.7\\
        &5&       $8\times8$                             & 4.4  & NA & NA \\
        &6&       $256\times256$                         & 12.1   & NA & NA\\\hline
    \end{tabular}
    \label{tab:pre-trained}
\end{table}

\begin{figure*}
    \centering
    \hspace{0.8cm}
    \begin{subfigure}[b]{0.8\textwidth}
        \includegraphics[width=\textwidth]{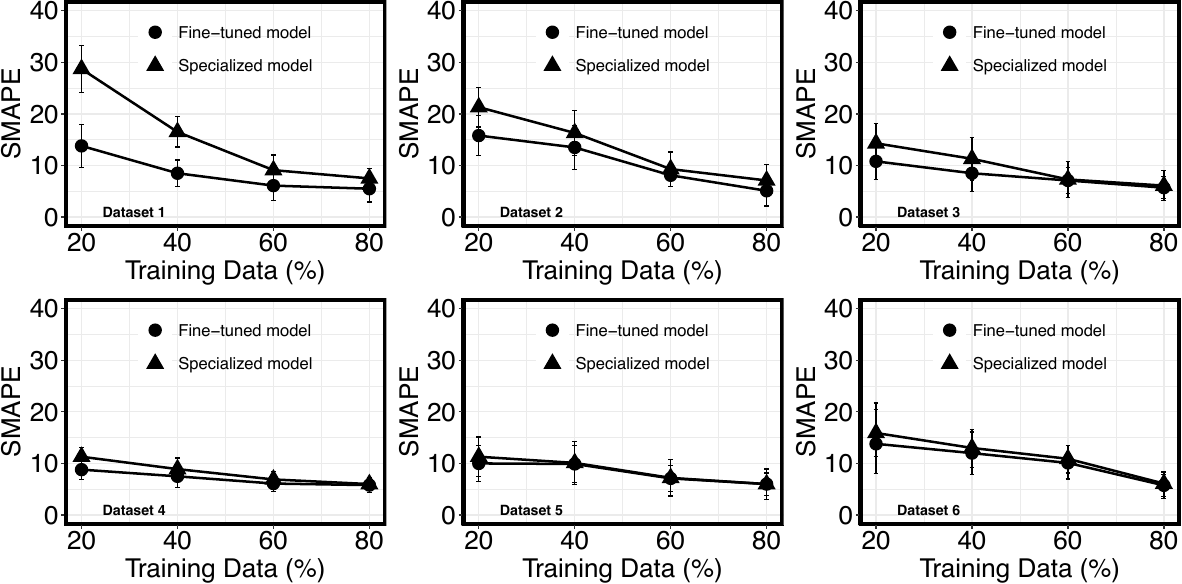}
        \caption{WiFi Dataset}
        \label{fig:training-wifi}%
    \end{subfigure}
    \newline
    \newline
    \newline
    \begin{subfigure}[b]{0.8\textwidth}
        \includegraphics[width=\textwidth]{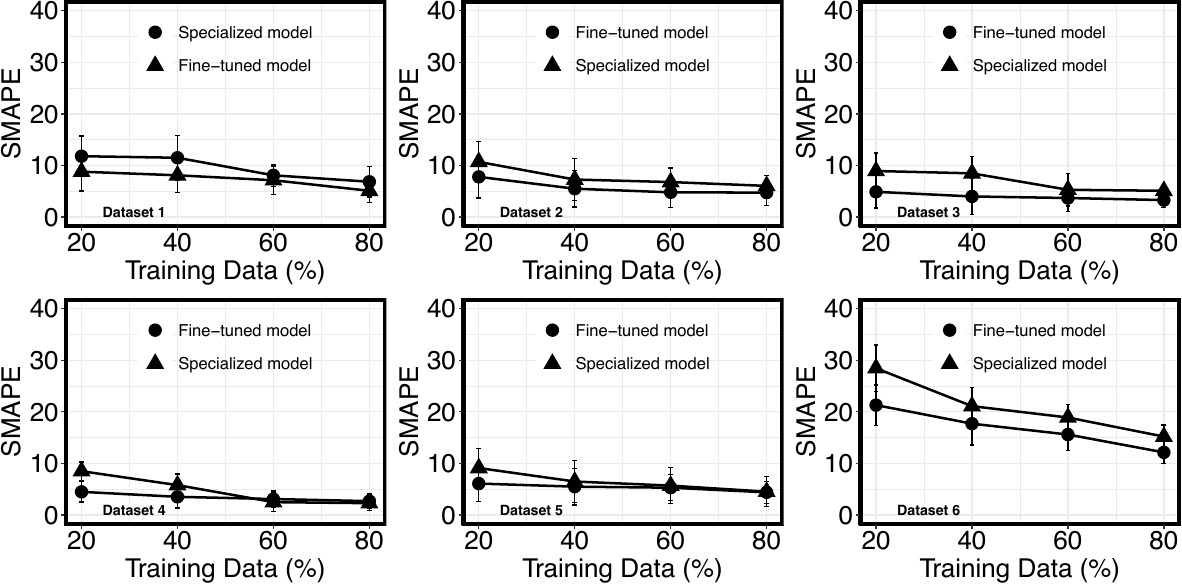}
        \caption{LTE Dataset}
        \label{fig:training-lte}%
    \end{subfigure}
    \caption{A comparison of the average SMAPE of the target model obtained by fine-tuning a source model and a specialized model trained from scratch for different target data sizes. The SMAPE is measured on the test dataset across
    10 out-of-sample validation runs. The error bars show a 95\% confidence interval. The source model for transfer learning is selected based on the minimum DTW measure between the source and the target dataset. And a single first LSTM layer was transferred.}
\end{figure*}

In this study, we select the pre-trained source model
using the minimum DTW measure between the source dataset and the target dataset.
In each fine-tuning, we transfer a certain number of layers beginning from the first layer
and then fine-tune only on the remaining higher layers. We vary the number of transferred layer parameter based on
the number of layers defined in the source model. For example, across all our source models
of WiFi and LTE datasets, only one source model consists of four LSTM layers (LTE dataset 5). In this case,
we perform fine-tuning by varying the number of transferred layer parameter from 1 to 3. Likewise, in the case of source models having three LSTM
layers, we vary the parameter from 1 to 2, and in the case of two-layered LSTM architectures, we set it to 1.  We repeat each
fine-tuning across 10 out-of-sample validation runs of the target dataset.

\tabref{tab:pre-trained} shows the average SMAPE values based on the number of transferred layers
for each WiFi and LTE dataset in the archive. In the case of a source model
with three layers, the average SMAPE of the fine-tuned model increases as more than one
layer is transferred. For a source model with four layers ($16\times16\times16\times16$), the SMAPE of fine-tuned model reduces until the
transfer of two layers but increases thereafter. Thus, increasing the transferred layers does not provide gains.
This can be well explained by the phenomenon confirmed in the transfer learning
studies in other domains~\cite{10.5555/2969033.2969197}.
The first layer in the LSTM architectures learns the generic features of the training dataset that are also applicable across other datasets
while the higher layers learn features that are more training task dataset-specific. Thus, when the higher layers are transferred in the
fine-tuning, it inhibits the architecture from learning the new target task dataset-specific features, thus negatively impacting the
predictive power of the target model.

\subsubsection{Target Training Data Size}
\label{sec:datasize_result}
Knowing how much target environment data to collect to be able to generate realistic synthetic traces is critical.
To answer this question, we vary the target environment data size used for fine-tuning the source model using
transfer learning. In this analysis, for each target dataset, we select a source model
using the minimum DTW measure. Based on our previous
analysis of the impact of the number of transferred layers on the target model performance, we transfer a
single layer which is the first LSTM layer of the source model and fine-tune it with the varying
target data sizes.

\figureref{fig:training-wifi}a (WiFi) and \figureref{fig:training-lte}b (LTE) illustrate the comparison of average SMAPE
between a target model obtained by fine-tuning a source model, and for specialized models trained from scratch at
different target data sizes. The error bars in the figures represent the 95\% confidence interval across
10 out-of-sample validation runs. 20\% target data size consists of 6000 RTT samples, while on the
other end 80\% consists of 24\,000 RTT samples.

The SMAPE values of both the models across all datasets and training data sizes
range from 28.7\% to 5.1\% for WiFi and 28.4\% to 2.7\% for LTE. In~\figureref{fig:training-wifi}a and \figureref{fig:training-lte}b,
one can see that the SMAPE of both models increase as the data size decreases.  This behavior is
persistent across both the WiFi and the LTE. Also, fine-tuning has the highest performance improvement
when the target data size is the smallest (20\%) for both the WiFi and the LTE. The SMAPE improvements for fine-tuning case ranges from
51.9\% to 3.9\% in WiFi, and 47\% to 25\% in LTE for 20\% data size. As the data size begins to increase, these gains
diminish. For the largest data size of 80\%, the gains are marginal (from 26\% to 0\% for WiFi and 35\% to 2\% for LTE);
in the few datasets, there is no gain at all. The results very well intuitively show that obtaining a good specialized
model requires a proper size of target environment data samples while fine-tuning is well done with short samples.

Another interesting aspect of the results is that for some datasets, the improvements in SMAPE between fine-tuned
(for 20\% data size) and specialized (for 40\% data size) models are also marginal. It shows that training a specialized
model from scratch using more data may not always bring in proportional performance improvements. Further, it also
highlights that by selecting a transfer learning approach, one can obtain a speed up by reducing down the target
environment data collection time.

The results could be used as a general guideline to estimate the minimum size of the target environment data
needed to build target models of a similar context with a certain accuracy. For example, an application tester interested
in building a cloud access time model for a vehicular LTE network connectivity in an urban environment can refer to our
LTE Dataset 1 and Dataset 2 in the results. From the results, one can infer with a certain degree of confidence
that a target environment data size of 6000 RTT samples, when fine-tuned with one of the source models, can generate
a target model whose average SMAPE can vary in the range of 4.7\% to 12.5\%.

Further, we quantify the speed improvement resulting from the usage of transfer learning as compared to training a specialized
model. We compute the speed-up factor of transfer-learning as the ratio of time required to train a specialized model from
scratch and the time required to fine-tune the source model. The speed-up factor is measured across 700 epochs.
A single first layer of the source model is transferred for fine-tuning. The training jobs run on \textit{n1-standard-4}
virtual machine (VM) type on GCP with 4 virtual CPUs and 15 GB of memory. The vCPU is implemented as a single hardware
hyper-thread on the Intel Xeon Scalable Processor (Skylake)\footnote{Intel Xeon Scalable Processor (Skylake) has a
2~GHz base frequency, 2.7~GHz All-core turbo frequency, and 3.5~GHz Single-core
max turbo frequency.}~\cite{gcpdocs}.

\tabref{tab:speed} shows the comparison of the speedup factor of fine-tuning and the corresponding average SMAPE improvement
percentage for all positive transfer cases across different target data sizes. The speed-up factor for a single transferred
layer is similar across different data sizes and ranges from 1.1 to 1.8 for the WiFi and LTE, while the SMAPE improvement percentage
varies from 1.3\% to 51.9\%. These results show that transfer
learning not only reduces the data collection time but also speeds up the model building time without adversely impacting
the target model accuracy. We conclude that carrying out transfer learning in an informed manner based on the dataset
similarity rarely hurts.

\begin{table}
    \centering
    \caption{Fine-tuning speedup on a \textit{n1-standard-4} virtual machine (VM) type on GCP.}
    \centering
    \begin{tabular}{|c|c|c|c|c|c|c|c|c|c|}
        \hline
        &      \textbf{Target}        & \multicolumn{4}{c}{\textbf{Fine-Tuning}} & \multicolumn{4}{|c|}{\textbf{SMAPE}}\\
        &      \textbf{Dataset}                     & \multicolumn{4}{c}{\textbf{ Speedup}} & \multicolumn{4}{|c|}{\textbf{Improvement (\%)}} \\
        \cline{3-10}
        &                              & 20 & 40 & 60 & 80 & 20 & 40 & 60 & 80 \\
        \cline{1-10}
        \multicolumn{1}{|c|}{\multirow{6}{*}{\parbox[t]{2mm}{\multirow{3}{*}{\rotatebox[origin=c]{90}{\textbf{WiFi}}}}}}         & 1 & 1.2  & 1.2  & 1.1 & 1.2 & 51.9  & 48.4   & 32.9   & 26.6\\
        \multicolumn{1}{|c|}{}& 2 & 1.2  & 1.2 & 1.5  & 1.2   & 25.8 & 17.1  & 12.9  & 15.1\\
        \multicolumn{1}{|c|}{}& 3 & 1.3   & 1.3    & 1.3   & 1.3    & 24.4  & 24.7   & 2.7   & 5.5\\
        \multicolumn{1}{|c|}{}& 4 & 1.1   & 1.1    & 1.1 & 1.3   & 22.1  & 15.7   & 11.5   & 3.3\\
        \multicolumn{1}{|c|}{}& 5 & 1.1  & 1.3  & 1.3 & 1.4  & 11.5 & 1.9 & 1.3 & 1.6\\
        \multicolumn{1}{|c|}{}& 6 & 1.3   & 1.3  & 1.2  & 1.3    & 13.2  & 7.6   & 7.3   & 5.7\\
        \cline{1-10}
        \multicolumn{1}{|c|}{\multirow{6}{*}{\parbox[t]{2mm}{\multirow{3}{*}{\rotatebox[origin=c]{90}{\textbf{LTE}}}}}}             & 1 & 1.2  & 1.5 & 1.8  & 1.4   & 34.0 & 25.4  & 29.5  & 11.8\\
        \multicolumn{1}{|c|}{}& 2 & 1.1  & 1.4 & 1.6  & 1.2   & 27.1 & 24.4  &  29.3   & 21.4\\
        \multicolumn{1}{|c|}{}& 3 & 1.1  & 1.1 & 1.3  & 1.2   & 52.8 & 45.1  &  30.1   & 39.2\\
        \multicolumn{1}{|c|}{}& 4 & 1.8   & 1.6 & 1.6  & 1.5   & 47.0  & 39.1 & 19.3 & 14.8\\
        \multicolumn{1}{|c|}{}& 5 & 1.3   & 1.2  & 1.2   & 1.1   & 33.0  & 15.3 & 7.0   & 2.8\\
        \multicolumn{1}{|c|}{}& 6 & 1.4  & 1.3 & 1.5  & 1.2   & 25.1     & 16.1  & 17.4  & 20.1\\\hline
    \end{tabular}
    \label{tab:speed}
\end{table}

\begin{figure*}
    \centering
    \includegraphics[width=0.8\textwidth]{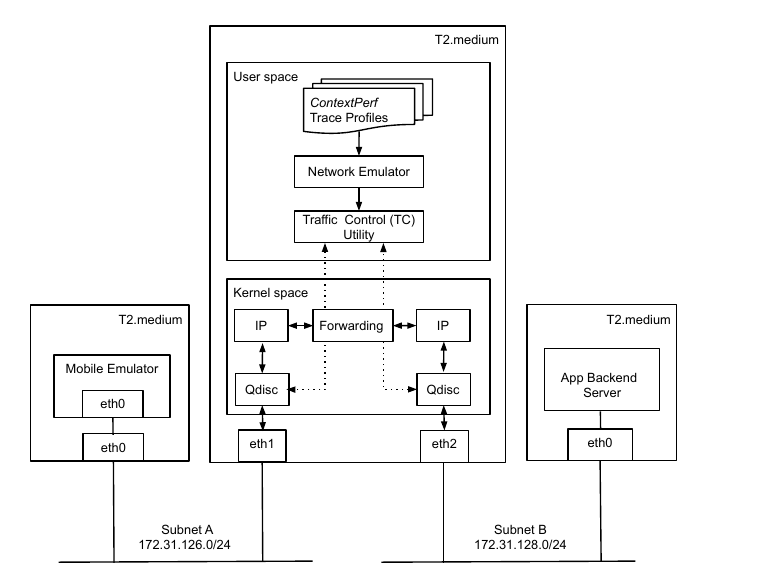}
    \caption{Experimental setup for mobile application testing: A mobile emulator, network emulator, and application backend server
    hosted on separate VMs within the cloud.\label{fig:emulatorsetup}}
\end{figure*}
\section{Usage of Transfer Learning in an emulation}
\label{sec:emulation}
In this section, our primary goal is to automate the process of selecting an appropriate source model, fine-tuning the source
model on the target dataset, and generating synthetic access times traces using this transfer learned model. Further, we aim
to integrate the synthetically generated traces within a network trace-based emulator so that it can be used in
real mobile application testing. In the following, we describe our implementation efforts in this direction in the
form of a tool called \textit{ContextPerf}. In the following, we present the internals of the network emulator, and its
integration with \textit{ContextPerf} generated traces. Finally, we evaluate the end network emulation accuracy
independently with real access time traces.

\begin{figure*}
    \centering
    \includegraphics[width=0.8\textwidth]{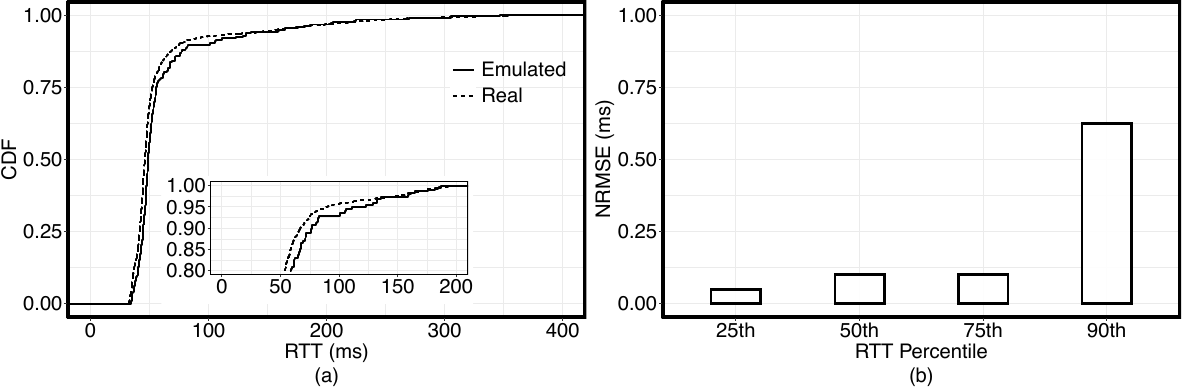}
    \caption{(a) Cumulative Distribution Function
    (CDF) comparing the RTT distribution of the real RTT trace fed to the network emulator and the RTT of the emulated network path for a single emulation run of 10~min. (b) Bar-plot of the Normalized Root Mean
    Squared Error (NRMSE) between the real RTT and the emulated network path RTT normalized using the standard deviation of the real RTT trace for 25th, 50th, 75th, and 90th percentiles. NRMSE is calculated across 30 emulations.\label{fig:emulatoracc}}
\end{figure*}

\subsection{Implementation}

\textit{ContextPerf} is a tool written in Python that provides end-to-end automation for the generation of synthetic
cloud access times traces. It takes as input a measurement sample trace file of a target environment. It standardizes
the measurement sample and selects a pre-trained source model from a model archive hosted in the cloud storage bucket.
The pre-trained source model is selected based on the minimum DTW distance to the target dataset. Next, based on the
input fine-tuning hyperparameters, it creates a transfer learning job in the AI platform service of GCP. Upon the
fine-tuning job completion, the model and the synthetic traces generated from it are made available as files in the
bucket. The synthetic trace files can then be used with any network emulator
for emulating cloud access times.

We integrate these synthetic traces with the mobile network emulator developed
by Akamai~\cite{akamai,DBLP:journals/corr/GoelSWLF17}.
The experimental setup for mobile application testing is shown in  \figureref{fig:emulatorsetup}. It includes the following
three components: the mobile emulator that runs the application binary under test, the network emulator which emulates the cloud access times
using~\textit{ContextPerf} generated synthetic traces, and the application cloud backend server. We deploy each of these components
on a separate Virtual Machine (VM) in the cloud. Going forward, we limit our discussion to the network emulator component,
its internal architecture, and integration with~\textit{ContextPerf} generated trace files.

As shown in  \figureref{fig:emulatorsetup}, the network emulator is hosted on a \textit{t2.medium} VM in the
AWS public cloud. The emulator requires a VM with two virtual network interfaces.
One virtual network interface belongs to the subnet of the mobile emulator (the application client) and the other belongs to
the subnet of the application backend server. The network emulator then uses
\textit{iptables}~\cite{iptables} rules with masquerading
to forward traffic received on the client subnet to the server subnet and vice versa. On the client and the server VMs,
the underlying routing tables are modified to route the network traffic through the network emulator VM.

The network emulator is based on the Linux Traffic Control (TC) \cite{tc}. TC
is a set of utility tools that enables
configuration and management of the kernel packet scheduler, and thus offers control over the packet traffic flow transmitted and received through
the queueing system (\textit{qdisc}) associated with the network interface. Using TC, one can introduce packet delays, change
the rate at which packets are transmitted and received, as well as drop and re-order the packets. Thus, one can emulate
the delay, bandwidth, packet loss, and re-ordering characteristics of
any network path. The network emulator is available as a bash script with an in-built set of fixed network profiles.

The emulator uses time-driven emulation that updates the delay parameter of the \textit{qdisc} after every fixed time
interval. This value is set to the time interval used to generate the synthetic traces. We set it to 500~ms.
The emulator internally adjusts this time interval to account for the time required to change the \textit{qdisc} using
the TC utility (4~ms).

We modified the source code of the emulator to change the delay parameter as per the synthetic traces
generated by \textit{ContextPerf}. Since \textit{ContextPerf} generates RTT synthetic traces, we divide them
to set the uplink and the downlink network path delay. We configure the network emulator to use symmetric
delays for the uplink and downlink paths.

We carried out the following modifications and extensions to the core functionality of the network emulator.
We added a configuration API support that enables the external feeding of the trace files as input to the
emulator. It supports setting the time interval for changing the emulation parameters and customizing
the ratio of the uplink to the downlink delay. We daemonized the network emulator component by implementing it as a \textit{systemd} service on the VM. The network
emulator service is then packaged as a VM image and made available as a Terraform~\cite{terraform} module using
the AWS provider.

\subsection{Evaluating Baseline Emulation Accuracy}

Next, we evaluate the baseline network emulator accuracy for the setup in  \figureref{fig:emulatorsetup} in an isolated manner
independent from the~\textit{ContextPerf}. Our goal is to find out how accurately the network emulator emulates the network
path using the cloud
access time traces fed to it. We analyze the RTT characteristics of the emulated network path between the mobile emulator
client and the application backend server in  \figureref{fig:emulatorsetup}.

In the experiment, we feed a real RTT trace to the network emulator and then measure the RTT on the emulated network
path. The RTT trace fed to the emulator consists of ICMP ping every 500~ms for a total
of 600 pings lasting 5~min. It is measured between a real Android phone and a VM in the AWS cloud.
We repeat the experiment for 30 different RTT traces measured over five Wifi and two LTE access
networks.

\figureref{fig:emulatoracc}a shows the Cumulative Distribution Function (CDF) comparison of a real RTT trace
fed as input to the network emulator and RTTs of the emulated path (\figureref{fig:emulatorsetup}). Each emulation run
lasts 600~s (10 min). As can be seen from the CDF in  \figureref{fig:emulatoracc}a, the 25th, 50th, 75th percentile values of the real input trace RTTs and those of the emulated path are the same while the 90th percentile
RTT values are within 3.3~ms of each other. We can see that the emulation also reproduces the long-tail nature of
the RTTs. We also plot the Normalized Root Mean Squared Error (NRMSE) between the real and emulated path RTTs
normalized using the standard deviation of the real RTT. The NRMSE is calculated for the 25th, 50th, 75th, and 90th percentile values across 30 different emulation runs, each lasting 10~min. An NRMSE value in the range 0 to 1 is a good value
where the emulation error is less than one standard deviation of the real RTT trace fed to the network emulator.
As can be seen from  \figureref{fig:emulatoracc}b all the NRMSE's are less than 1~ms with the one for the 90th percentile being the highest, 0.62~ms. From the above results, we conclude that the network emulator can accurately emulate
the cloud access times between the mobile and the cloud backend including its long tail nature.

\section{Case Studies}
\label{sec:casestudies}

In this section, we demonstrate the accuracy of our methodology compared to the popularly used normal distribution
access time
models~\cite{DBLP:conf/mobisys/ManweilerAZCB11,10.1145/1814433.1814452,10.1145/2348543.2348583}.
For this purpose, we evaluate QoE metrics of real mobile applications using the developed emulation environment under
our fine-tuned models and normal distribution models.

\subsection{Instagram}

In the first case study, we focus on Instagram~\cite{insta}, a popular social
media mobile application extensively used
for sharing photos and videos. We consider a
hypothetical scenario in which an Instagram tester is interested in profiling
the impact of cloud access time on the application level latency for the common user action of sharing a photo from the user's device. The study is to be carried out for the LTE network of a specific MNO using a stationary device in an indoor home environment. The photo is
uploaded on the cloud backend and shown to users in their feed. For this action, the application photo sharing
latency is the Above The Fold Time(AFT) latency and measured as the time from when the share button is pressed by the
user to the time when the posted photo's last pixel appears on the user's feed. In order to measure the AFT
latency, we capture video of the mobile screen labeled with the time clock while performing the action and
calculate the time delta using the method proposed by Brutlag \etal~\cite{aft}.

The measurement setup for the performance evaluation is the same as in  \figureref{fig:emulatorsetup}. We have two types
of emulators, the Android mobile emulator, and the network emulator. The Android mobile emulator is deployed in our context
simulation testbed~\cite{DBLP:conf/percom/RegeHW17,10.1145/2494091.2499578} for
mobile application testing based on AWS. It has
Instagram application installed on it.
The mobile and network emulators are deployed on separate VM instances within the testbed (\textit{t2.medium} type with 2
virtual CPU cores, and 4.0 GB RAM).

Now following our methodology, we collect 2500 RTT samples to the Instagram backend over the LTE network of this specific MNO using a stationary mobile device in an indoor home environment. The 2500 RTT samples are shorter than the minimum
data size of 6000 and approximately an order of magnitude smaller than the largest data size of 24\,000 considered
in our study(see  \figureref{fig:training-lte}b). We use the~\textit{ContextPing} application running on a
real Android mobile phone to collect RTT data.
A configuration of sending an ICMP ping for every 500~ms is used for RTT data collection. The measurement
lasts for a time duration of approximately 21~min. The task of collecting such RTT samples in the real scenarios could be crowd-sourced.

The target environment model is built by feeding the RTT samples to \textit{ContexPerf}.~\textit{ContexPerf}
pre-processes the target environment data of 2500 RTT samples by standardizing them, and splitting them in 80:20 ratio
for fine-tuning and testing, respectively. It selects a source model based on the minimum DTW measure between the source
and the target environment dataset and then fine-tunes it by transferring a single layer from the source models. The SMAPE of
the built target model is 14.3\%. The median value of the target environment dataset is used as input to the target
model to generate synthetic RTT traces from it, one step at a time for 2500 samples. These synthetic traces are then fed to the network emulator.
For the performance evaluation, we also set the bandwidth and packet loss of the network in the emulation that is measured separately (also 2500 samples) using Iperf~\cite{iperf}. As Iperf requires
access to the cloud backend, we
carried out the bandwidth and packet loss measurements separately to our dummy server in AWS which was hosted in the same region as the Instagram cloud backend.
For a baseline comparison, we also carry out emulation with access time traces generated from the normal distribution model. The mean and the standard deviation parameters are obtained by fitting a normal distribution to the target environment RTT data.

\begin{table}[!htp]
    \centering
    \caption{Statistical characteristics of access time traces of the LTE network --- Real vs. Normal distribution vs. Synthetic}\label{tab:statchar}
    \centering
    \begin{tabular}{|c|c|c|c|}
        \hline
        \multirow{2}{*}{\textbf{Statistic}}& \textbf{Real LTE} & \textbf{Normal Distribution}& \textbf{Synthetic}\\
        &      \textbf{(ms)}        & \textbf{(ms)}                         & \textbf{(ms)} \\\hline
        0.01\textsuperscript{th} Percentile  &   36.8     &  33.7   &   35.3     \\
        50\textsuperscript{th} Percentile    &   43.0     &  43.5   &   43.9    \\
        99.99\textsuperscript{th} Percentile &   58.1     &  52.8   &   60.1    \\
        Mean               &   43.4      &  43.5   &   44.5   \\
        Standard Deviation &   4.2       &  4.1    &   5.2   \\\hline
    \end{tabular}
\end{table}

A comparison of the statistical characteristics of the access time traces generated from our approach, the normal distribution,
and the real LTE network is shown in \tabref{tab:statchar}. From the results, one can see that both the synthetic traces
obtained from our approach and the normal distribution traces have a similar statistical mean, median as that on the
real LTE network. However, the synthetic traces perform better at extreme percentiles.

Next, we analyze the impact of these access time traces on the Instagram application AFT photo sharing latency.
We wrote a simple script using Appium~\cite{appium} to automate photo sharing
action on Instagram. In each photo-sharing action experiment, a photo from the library is selected and shared on the profile. The screen of the
emulator is recorded in the background using the Android \textit{screencapture} tool. We repeat the experiment for
50 runs with 2~min intervals.

\begin{figure*}
    \centering
    \includegraphics[width=0.7\textwidth]{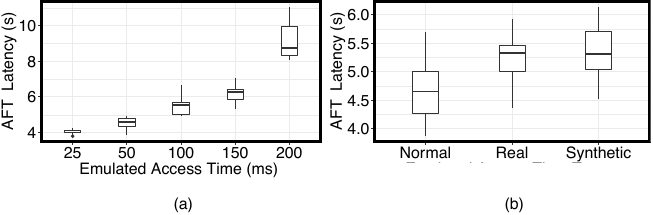}
    \caption{Boxplot of \textit{Instagram} AFT application photo sharing latency (seconds) measured with (a) different fixed emulated access times
    (b) emulation using the synthetic and normally distributed access time traces, and on an equivalent real LTE network.\label{fig:case-study-instagram}}
\end{figure*}

In order to study the impact of access time on the AFT latency, we first ran the experiments under
the emulation of different fixed access times. The results (\figureref{fig:case-study-instagram}a) show that the AFT latencies
increase with an increase in the underlying access times. For the access time of 200~ms, the AFT latency can
get as high as 10 s, thus leading to a significant degradation in QoE. These results indicate that access time is part of
the critical path. Our findings concur with the analysis of the post sharing action of the Facebook mobile application
made by Chen \etal~\cite{10.1145/2663716.2663726}.

Next, we repeat the above experiment by varying the underlying network emulation, first using our synthetic traces, and
then with normally distributed traces. Finally, we also run the experiment on the real LTE network in the same home environment.
\figureref{fig:case-study-instagram}b shows the comparison of Instagram application AFT latencies measured under the three scenarios.
The box in the figure is defined by the 25th, 50th, 75th percentiles.

In~\figureref{fig:case-study-instagram}b, one can see the variation in the measured AFT latencies across the two models
and the real LTE network. The AFT latency on the real network varies from 4.3 s to 5.9 s with mean and median values being
5.2 s and 5.3 s, respectively. In the case of an emulated network with normally distributed access times, the AFT latencies
are symmetrically distributed around the mean and the median. Both the mean and median values are 4.6 s.
In comparison to the real LTE network, the mean and the median is lower by 13\%. The maximum is 5.6 s which is lower by 8\%, and the minimum is 3.8 s which is lower by 15.5\%. In the case of application AFT latencies
measured under the network emulation using our generated synthetic traces, both the mean and median values are 5.29 s which
translates into an error of 2.0\% and 0.4\% compared to the real LTE network. The maximum value (6.1 s) is lower by 3.3\%,
while the minimum value (4.5 s) is lower by 3.6\%. These error values indicate that compared to the popular normal distribution,
the synthetic traces obtained using our fine-tuned model accurately estimate the AFT latencies across different  quantiles.
\subsection{Conversations}

Our second case study is an Android chat messenger mobile application.
\textit{Conversations}~\cite{conversations} is a
mobile application that uses an Extensible Messaging and Presence Protocol (XMPP) for message communication. XMPP is an
XML-based communication protocol for message-oriented middleware used by many popular chat messenger applications.
\textit{Conversations} application client supports sending and receiving images, voice messages, and files with OpenPGP
end-to-end message encryption. We use an XMPP based Ejabberd~\cite{ejabberd}
server as our backend for the~\textit{Conversations}
application.

We suppose that an application tester is interested in analyzing the QoE metric called message delivery receipt
latency for the \textit{Conversations} application in various target environments. Message delivery receipt latency
is measured as the time duration beginning from when the sender sends a message to the time when the sender
receives an acknowledgment receipt for the successful delivery of that message to the recipient and displayed to the
sender on the device. The delivery latencies are to be measured for a combination of different access network technologies and
end-user context environment in \tabref{tab:scenario}.

\begin{figure*}
    \centering
    \includegraphics[width=0.7\textwidth]{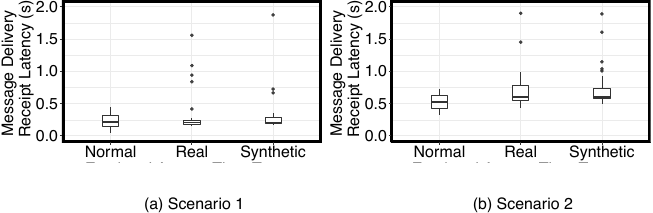}
    \caption{Boxplot of \textit{Conversations} application message delivery receipt latency (seconds)
    measured for different scenarios in \tabref{tab:scenario} with emulation
    using the synthetic and normally distributed access time traces, and on equivalent real networks.\label{fig:case-study}}
\end{figure*}

The measurement setup for the performance evaluation is similar to the one used in  \figureref{fig:emulatorsetup}.
The Ejabberd backend server is hosted in AWS in the EU-Ireland region. We use two separate Android emulators for a
sender and a receiver with \textit{Conversations} applications installed on them. Both Android emulators run on a separate
VM instance (\textit{t2.medium}) in our AWS based testbed. Further, to avoid the network traffic interference from the two mobile
emulators, we use separate VM instances for the network emulators (also a \textit{t2.medium}). Thus, in total,
we have the following five components running on five VMs in our experimental setup: the two mobile emulators, and their two network emulator
instances, and a Ejabberd backend server.

\begin{table}[!htbp]
    \centering
    \caption{Scenarios for measuring message delivery receipt latency for~\textit{Conversations} application}\label{tab:scenario}
    \centering
    \begin{tabular}{|c|c|c|}
        \hline
        \textbf{Scenario}          &  \textbf{Sender Environment}    &  \textbf{Receiver Environment}\\\hline
        1          &  LTE Mobility Vehicle (Berlin) & WiFi Home (Berlin)\\
        2          &  WiFi Home (Mumbai) & WiFi Home (Berlin) \\\hline
    \end{tabular}
\end{table}

We collect a short sample data of access times of each target environment to build its model with
fine-tuning. The appropriate sample data size necessary for fine-tuning is determined by referring to the results of our earlier studies for a similar target environment (\sectionref{sec:datasize_result},  \figureref{fig:training-wifi}a
and  \figureref{fig:training-lte}b). From the results, a fine-tuned model with a data size of 6000 RTT samples has a
SMAPE in the range of 8.0\% to 21.7\% for WiFi networks(Home and Office --- Dataset 3, 4, 5, and 6) and 4.7\% to 12.5\% for
the LTE mobility vehicle scenario (Dataset 1 and 2). On the other end, a fine-tuned model with 24\,000 RTT samples leads to
a SMAPE in the range 3.0\% to 12.7\% for WiFi networks and 2.2\% to 7.2\% for the LTE networks. Based on these results, we decide to collect 2500 RTT samples for scenarios in \tabref{tab:scenario}. This sample size is shorter than the minimum sample size of 6000 used in our study. The measurements for the sender and the receiver environments in each scenario are carried out at the same time.

The target environment models are built by feeding the RTT samples to \textit{ContexPerf}. The SMAPE of the obtained fine-tuned target models are 17.2\% (LTE mobility vehicle --- Berlin), 26.7\% (WiFi Home --- Berlin), and 24.5\% (WiFi Home --- Mumbai), 20.6\% (WiFi Home --- Berlin). The median value of the target environment dataset is used as input to the respective target model to generate its synthetic RTT trace for 2500 samples.
The synthetic traces are fed to the network emulators together with the real measured bandwidth and
packet loss traces. As in the earlier case study, the traces generated from the normal distribution model are used for baseline comparison.

The experiment to measure the message delivery receipt latency of~\textit{Conversations} application consists of
sending a fixed-size 2~kB message from the sender application. The source code of the application is instrumented to log the
timestamps of the message sent and acknowledgment receipt delivery events. The difference in the timestamps is
calculated to obtain the message delivery receipt latency. For each of the emulation scenarios, the experiment is
repeated for 50 runs. To compare the emulated message delivery receipt latencies of~\textit{Conversations}
application with the real world, the experiment is repeated for 50 runs on the real access networks in the end-user context
scenarios (\tabref{tab:scenario}). In order to minimize the impact of time and diurnal patterns on the network performance,
these experiments were carried out sequentially immediately after the RTT samples of the target environment were
collected as explained above.

\figureref{fig:case-study}a and~b show the box plots of the message delivery receipt latencies
for the \textit{Conversations} application in the two scenarios. The box in the figure is defined by the 25th
and the 75th percentiles while the middle line represents the median 50th percentile.
It is evident from both the figures that the delivery receipt latency distribution on the real network is positively skewed.
For scenario 1 (\figureref{fig:case-study}a) that consists of a message sender on an LTE network in a mobile vehicle, and a stationary receiver
on a WiFi home network, both in the Berlin metropolitan, the mean message delivery receipt latencies are 0.28 s (real network),
0.27 s (synthetic trace), and 0.23 s (normal) while the median latencies are 0.20 s, 0.20 s, and 0.21 s, respectively. It can be seen
that the mean and the median delivery receipt latencies of the \textit{Conversations} estimated by the normal distributed access
times traces and our synthetic access time traces are quite similar. And in comparison to real networks, both models accurately estimate the mean and
median latencies. On the other end, the estimated maximum delivery latencies are 1.34 s (real network), 1.33 s
(synthetic trace), and 0.42 s (normal), and the minimum latencies are 0.15 s, 0.18 s, and 0.05 s, respectively. The accuracy gap between
the two models widens at extreme percentiles. A normally distributed access time model produces symmetric delivery receipt
latencies. Furthermore, in scenario 1, it significantly underestimates the maximum and the minimum receipt delivery latencies by 72.2\% and 84.4\%, respectively. In scenario 2, similar behavior is observed with values underestimated by 61\% and 24.4\%,
respectively. In contrast, our synthetic traces overestimate latencies by 20\% and 21.3\% (scenario 1) and 0.5\% and \tmin13.2\% (scenario 2).
The results show that the emulated network environment using synthetic traces obtained from fine-tuned models can accurately reproduce
message delivery receipt latencies including the outlier values observed on the real networks.

\subsection{Summary}

Our case studies quantify the impact of access time model selection on the QoE metrics of two different categories of mobile applications
viz. social media and chat messenger. Our results show that in both the applications, choosing the popularly used simple normal distribution
access time model can provide an accurate estimation of the mean and median QoE metrics. However, the normal distribution model fares poorly at
the extreme percentiles, providing conservative estimates of the QoE metrics. This can be attributed to the fact that the underlying
access times obtained from a normal distribution model are symmetric along the mean access time value. In the light of earlier studies
by Amazon~\cite{10.1145/1281192.1281295} and Google~\cite{marissa}, these
differences are quite significant. Using the popular normal
distribution model for access time emulation can lead a tester to obtain overall better QoE metric estimates (message delivery receipt latency, AFT latency)
and thereby falsely interpret it as superior application performance. Our case-studies also successfully demonstrate the superiority
of synthetic access time traces obtained from target environment-specific fine-tuned models in mobile application testing, and the
ease with which they can be realistically generated with a relatively short amount of measurement sample data.

\section{Related Work}
\label{sec:related_work}

\textbf{Cloud Access Time Models ---} Modeling cloud access times as end-to-end delays have been extensively
studied in the
literature~\cite{1007528,4146998,1673218,4381513,OHSAKI20011027,Paxson:1997:EIP:263109.263155,Sanaga:2009:MEI:1558977.1558991,8717870}.
A time series of end-to-end delay is a sequence of data points measured at
certain time intervals.
Therefore, the modeling problem has been formulated as a prediction problem to estimate the next data points in
the time series. We refer the readers to Yang \etal~\cite{1295650} for a
detailed overview of various delay prediction techniques
and discuss some of them briefly in the following. The widely used analytical models for end-to-end delay include
Hidden Markov Models (HMM)~\cite{8717870} and Autoregressive (AR)
models~\cite{OHSAKI20011027}.
From the empirical perspective, Rao~\cite{1258117} presents an approach for
delay prediction based on regression. On using
the machine learning approach, Belhaj \etal~\cite{belhaj09}
and Parlos \etal~\cite{1007528} have
used Recurrent Neural Networks (RNN) for modeling the delay dynamics. Trevisan
\etal~\cite{TREVISAN2020107289} apply Kernel
Density Estimation (KDE) to model mobile network characteristics. Yang
\etal~\cite{4146998} propose a multiple
model approach that predicts delay using the combination of estimates from various models treated as a bank of filters.
Bui \etal~\cite{4381513} use wavelet transforms in combination with a recurrent
multilayer perceptron neural net for
long-horizon end-to-end delay prediction. There have been several measurement-based studies that have tried to characterize
the end-to-end delay over different types of access networks. For instance,
Manweiler \etal~\cite{10.1145/1999995.2000003}
and Huang \etal~\cite{10.1145/1999995.2000003} have characterized the latency
over 3G networks and found them to be
normally distributed. All the above prediction techniques are largely data-driven, and using them for other targeted environments
requires a long and laborious data collection. For instance, Belhaj
\etal~\cite{belhaj09} and Parlos \etal~\cite{1007528},
Yang, \etal~\cite{4146998} collect data in the order of magnitude of hours
lasting an entire day.~\cite{TREVISAN2020107289} ran
a large-scale data collection campaign to obtain cloud access time data. This is not feasible for a tester of a single application
or few applications. Unlike these studies, in our work, we focus on the problem of using a relatively shorter sample of the
targeted environment in the order of magnitude of minutes to fine-tune a selected pre-trained source model.

\textbf{Network Emulation Tools ---}
Network emulation tools are also being used in analyzing the QoE of mobile applications. Network emulation has been well-studied
in the literature. Linux Traffic Control (TC)~\cite{tc} is a set of utilities
that enable controlling the packet traffic flow through the queueing system
associated with the network interface. NetEm utility in TC has been extensively
used for emulating various network conditions. Another similar tool is the
network link emulator Dummynet~\cite{Rizzo97dummynet:a} and Wide Area Network
(WAN) emulator NIST Net~\cite{10.1145/956993.957007}. These tools provide
simple traffic shaping policies that can emulate fixed latencies as well as
latencies belonging to well-known parametrized distributions like the normal
distribution. The standard parametrized distributions like normal distribution
are not capable of accurately emulating the time-based
variability of the access times. In our work, we rely on TC for trace-driven emulation with~\textit{ContextPerf} generated synthetic access
time traces. As opposed to using these tools directly for network emulation in QoE measurement, using our approach provides access network technology-specific and target environment-specific cloud access time emulation that preserves time-based variation.
Furthermore, several other network emulation tools have been built based on traffic shaping algorithms viz.
Augmented Traffic Control (ATC)~\cite{atc} from Facebook, Network Link
Conditioner~\cite{nlc} for MacOS, Google Chrome Devtools~\cite{devtools},
Android Emulator~\cite{androidemu} network emulator, etc. A feature improvement
in these tools has been the addition of
a network profile collection containing selected network access time traces. The traces are organized in various categories viz. Poor, Good, Lossy
based on their mean and standard deviation statistics. As discussed earlier in  \sectionref{sec:intro}, such tools
have trace profiles within a defined scope which offer limited access time variation dynamics. In contrast, the traces obtained
from our access time modeling approach are more representative of a wider population of target environments.

\textbf{Mobile Application QoE Measurement Tools ---}
In this scope, there has been extensive research in building mobile application QoE measurement tools.
Chen \etal~\cite{10.1145/2663716.2663726} present a tool called QoE Doctor
that uses UI automation techniques
to generate user sessions to measure mobile application QoE. It also supports cross-layer analysis of mobile applications
across application, transport, network, and cellular radio link layers. Another
QoE analysis tool WebLAR~\cite{DBLP:conf/pam/AsreseWBLAO19}
measures the web QoS metrics such as TCP connection time, and Time To First Byte (TTBF) and web QoE metrics like Above The Fold (AFT)
latencies and Page Load Times (PLT). WebPageTest~\cite{webpagetest} is another
such service to measure the QoE of web applications
on mobile devices. For video streaming applications like Youtube, Jimenez
\etal~\cite{8721061} uses the network packet-level data to
model the QoE of video sessions. YoMoApp by Wamser \etal~\cite{7194076}
measures the Youtube video session QoE based on the player state/events,
buffering, and video quality level data. All these tools focus on certain
categories of applications. In order to measure the impact of network
performance on the QoE metrics, these tools either require exposing the
application to the real network conditions or use some network emulation tool, as discussed earlier. As we demonstrated in
\sectionref{sec:casestudies}, traces generated with~\textit{ContextPerf} can be
effectively used to measure the QoE metrics across different categories of
mobile applications. On the contrary, our approach complements these QoE
measurement tools; it can be integrated with them to offer improved support for
cloud access time emulation.
\section{Future Research Direction}
\label{sec:discussion}

In this section, we discuss overarching issues related to our work and future research direction.

\textbf{Applicability to other network QoS metrics --- }While our work focuses on cloud access times, it is not the only network QoS metric that
influences the application performance and its QoE. Certain categories of applications are also highly influenced by network bandwidth
and packet loss. For example, video streaming applications like Youtube, audio streaming-based services like Spotify, etc. The applicability of our methodology
to build bandwidth and packet loss models need to be studied and analyzed.

\textbf{Widen the scope of target environments --- }In this work, we used a limited set of end-user contexts for building
fine-tuned models. We want to evaluate~\textit{ContextPerf's} capability under a broader range of scenarios. We are working on expanding
our scope to a wider range of target environments including the effects of control plane latencies, MNOs, geographical regions,
and times of the day, etc.

\textbf{Applicability to next generation 5G networks and use-cases --- }
The next generation 5G cellular networks aim to offer improved network QoS with 1000 times greater throughput
improvement, 100 billion connections at a massive scale that include Machine To Machine (M2M) communication, and a
close to zero latency ($<$1
ms)~\cite{DBLP:journals/comsur/AgiwalRS16,7169508}. With the introduction of
the Enhanced Mobile Broadband (eMBB), Massive Machine Type Communications (mMTC), and Ultra Reliable and Low Latency Communications (uRLLC) services
in 5G, there would be application use-cases in heterogeneous domains such as factory automation, connected vehicles,
robotics, virtual reality, health care, smart city, etc. These application use-cases would have their own specific
network QoS requirements. 5G stresses greater flexibility for the core network through software virtualization,
thus enabling easier instantiation of the core services and simplifying radio-resource allocation without compromising
on the network stability. It will also enable the different services verticals through resource sharing
using network slicing~\cite{7962821,7926923}. The impact of such network
slicing decisions on the network QoS and
and the application QoE within the vertical would need to be studied. Understanding the QoE of the use-cases will be a key
enabler of future 5G business cases. We envision an increasingly common need for target environment context
driven testing in each of these service verticals where our methodology could be applied. We aim to
build transfer learned 5G network QoS models, study their accuracy, and apply them to measure QoE.

\textbf{\textit{ContextPerf} automation tool release --- }We plan to make the ~\textit{ContextPerf} tool available to the
mobile application development and the testing community to enable testers to evaluate their mobile applications QoE for a wide range of target environments.
In this direction, we are working on developing and deploying ~\textit{ContextPerf} in the cloud and offer it in
form of a Network Testing as a Service (NTaaS). We are targeting wider testing scenarios where testers can run cloud
access time tests as a part of their periodic application release cycle.

\textbf{Real world in-context trace collection --- }While our work improves on shortening the real-world cloud-access time
traces needed to build target environment-specific models, we do recognize that sometimes access time traces might be
difficult to collect. This problem can be solved to a certain extent by providing the testers with a mobile application
like~\textit{ContextPing} that handles network trace collection parameterization related to granularity, scale, duration, etc.
while the labeling of context data is handled by the tester. Crowdsourcing offers a promising approach to scale up trace
collection. However, the collection of network trace data labeled with precise target environment context information
without any user intervention comes with its own set of challenges. We leave it as a part of future work.

\textbf{Applicability to other platforms -- Emulators and real mobile devices --- }Currently, we implemented the access time emulation by integrating
it in Traffic Control (TC) utility available on Linux. This limits ~\textit{ContextPerf's} ability to run seamlessly on mobile application
developer and tester workstation as a part of their development workflow. Our design enables us to integrate it with other mobile emulator
platforms like Android Emulator, and iOS. Furthermore, the applicability of our cloud access time emulation using a real mobile device and
its trade-offs with the mobile emulator needs to be studied and verified. We also consider it a part of future work.

\section{Conclusions}
\label{sec:conclusions}
In this paper, we addressed the problem of generating realistic synthetic cloud access time traces for various contexts.
We developed a methodology to model cloud access times with a Long Short Term Memory (LSTM) neural net
using a transfer learning framework. The methodology requires a relatively short sample of cloud access time trace
from a target environment to fine-tune an existing source model. To this effect, we prototyped an automation tool
called~\textit{ContextPerf}, which streamlines the process of building a fine-tuned model and generating synthetic traces
of context-based cloud access times from it. In real mobile application testing case-studies, we have compared the impact
of different types of cloud access modeling on the application QoE. We observe that usage of the popular normal
distribution access time model, as well as, of our fine-tuned model result in an accurate estimation of the mean and
median QoE metrics (AFT application latency, and message delivery receipt latency). The normal distribution
access time models, however, perform poorly with respect to the estimation of extreme percentiles of QoE. For example, in
the case of the popular \textit{Instagram} application, the maximum value of the AFT application latency for photo sharing
action, estimated by the normal distribution model has an error of 8\%, while our fine-tuned models estimate it with an error
of 3.3\%. Likewise, for the chat messenger application \textit{Conversations}, the maximum value of the QoE metric message
delivery receipt latency has an estimation error of 72.2\%, while our fine-tuned models estimate it with an error of only 20\%.
The two concrete case studies demonstrate that using our fine-tuned models result in a better assessment of QoE quantiles.

\bibliographystyle{IEEEtran}
\IEEEtriggeratref{86}
\bibliography{sigproc}

\end{document}